\documentclass[11pt]{article}

\usepackage[T1,T2A]{fontenc}
\usepackage[cp1251]{inputenc}
\usepackage{textcomp}
\usepackage[centertags]{amsmath}
\usepackage[mediummath]{nccmath}
\usepackage{amsfonts}
\usepackage{amssymb}
\usepackage[pdftex]{hyperref}
\usepackage{subcaption}
\usepackage{graphicx}
\usepackage[numbers,sort&compress]{natbib}

\usepackage{paperinitial}

\paperinitialization{15mm}{15mm}{15mm}{15mm}{2pt}{10pt}

\DeclareMathOperator{\re}{Re}
\DeclareMathOperator{\im}{Im}
\DeclareMathOperator{\sgn}{sgn}

\DeclareMathOperator{\rot}{rot}

\newcommand{\lan}{\langle}
\newcommand{\ran}{\rangle}
\newcommand{\bs}{\boldsymbol}

\newcommand{\e}{\varepsilon}
\newcommand{\vf}{\varphi}

\newcommand{\s}{\sigma}
\newcommand{\Si}{\Sigma}
\newcommand{\al}{\alpha}
\newcommand{\be}{\beta}

\newcommand{\de}{\delta}

\newcommand{\la}{\lambda}

\newcommand{\spx}{\mathbf{x}}
\newcommand{\spy}{\mathbf{y}}

\newcommand{\spk}{\mathbf{k}}

\newcommand{\spe}{\mathbf{e}}

\newcommand{\R}{\mathbb{R}}

\begin{document}
\allowdisplaybreaks[4]
\frenchspacing

\title{{\Large\textbf{Exploring spatial dispersion in helical wired media: \\ An effective field theory approach}}}

\date{}

\author{P.O. Kazinski\thanks{E-mail: \texttt{kpo@phys.tsu.ru}}\;\, and P.S. Korolev\thanks{E-mail: \texttt{kizorph.d@gmail.com}}\\[0.5em]
{\normalsize Physics Faculty, Tomsk State University, Tomsk 634050, Russia}
}

\maketitle

\begin{abstract}

The propagation of electromagnetic waves in helical media with spatial dispersion is investigated. The general form of the permittivity tensor with spatial dispersion obeying the helical symmetry is derived. Its particular form describing the medium made of conducting spiral wires with pitch $2\pi/|q|$ and chirality $\sgn(q)$ is studied in detail. The solution of the corresponding Maxwell equations is obtained in the paraxial limit. The dispersion law of the electromagnetic field modes, their polarization, and the integral curves of the Poynting vector are analyzed. The dispersion law of photons in such a medium possesses polarization dependent forbidden bands. The widths of these gaps and their positions are tunable in a wide range of energies. If the helix angle $\al$ is not close to $\pi/2$ and the plasma frequency $\omega_p\ll|q|$, then there are two chiral forbidden bands. The energies of one chiral forbidden band are near the plasma frequency $\omega_p$ and the width of this gap is of order $|q|$. The other chiral forbidden band is narrow and is located near the photon energy $|q|$. In the case $\al\approx\pi/2$, the first chiral forbidden band becomes a total forbidden band. If, additionally, the plasma frequency $\omega_p\gg|q|$, then the second forbidden band turns into a wide polarization dependent forbidden band. For the energies belonging to this interval the photons with only one linear polarization are transmitted through the medium and the polarization plane of transmitted photons is rotated. In the nonparaxial regime, the solution of the Maxwell equations is obtained in the shortwave approximation. The dispersion law of the electromagnetic field modes, their polarization, and the integral curves of the Poynting vector are found. Scattering of the electromagnetic waves by a slab made of the helical wired medium is considered.

\end{abstract}

\section{Introduction}

Chiral media both natural and artificial find their applications in a control of the properties of electromagnetic fields \cite{Belyakov1982,deGennes1993,Lakhtakia2005,Yang2006,Yang2012,Barboza2015,Kashke2016,Belyakov2019,Vetrov2020,Li2021}. One of the particular cases of the chiral media is represented by a helical medium that is the medium with permittivity tensor possessing the helical symmetry. Among the helical media are the cholesteric liquid crystals, the $C^*$-smectics \cite{Li2021,Vetrov2020,Belyakov2019,Barboza2015,Yang2006,deGennes1993,Belyakov1982}, and the metamaterials composed of dielectric \cite{Lakhtakia2005,Lee2005,Hung2011,Karakasoglu2018,Lei2018,Kilic2021,Cheng2022} or \cite{Kashke2016,Yang2012,Gansel2012,Liao2014,Li2015,Kashke2015,Jen2015,Venkataramanababu2018} conducting spirals. It is known that the electromagnetic waves in such media possess a chiral forbidden band \cite{Belyakov1982,deGennes1993,Lakhtakia2005,Yang2006,Yang2012,Barboza2015,Kashke2016,Belyakov2019,Vetrov2020,Li2021,Lakhtakia1995,Gevorgyan2021,KK2022,Wu2010,LiWongChan2015}. This property makes it possible to use these media as filters or converters of the circular polarization of the electromagnetic waves propagating along the helical axis of the medium. The polarization of the electromagnetic waves passed through the helical medium in a crosswise direction perpendicular to the helical axis also has certain peculiarities depending on the parameters of the medium. This allows one to construct broadband phase plates \cite{Wu2011,Balmakov2013,LiWongChan2015} employing the helical media. The presence of conducting structures in a helical medium enhances the polarization properties of the material. In the framework of electrodynamics of continuous media, conductivity of matter gives rise to a considerable spatial dispersion \cite{LandLifshECM}. In the present paper, we will study the helical media with strong spatial dispersion.

As a rule, the theoretical description of properties of the electromagnetic waves in conducting helical media relies on numerical simulations. In some particular cases where the helices constitute a crystal, it is possible to construct the semianalytical models \cite{Wu2010,Wu2011,LiWongChan2015} the ultimate analysis of which is also performed numerically. Beside their high labor and time consumption, such approaches do not provide an insight of the physical processes developing in the helical media. That does not alow one to predict the main characteristics of electromagnetic fields for the different parameters of such media without performing the laborious calculations. One of the methods used to describe, without much effort, the behavior of the longwave electromagnetic waves in the structured media is the averaging of microscopic Maxwell equations. It leads to the effective Maxwell equations with the permittivity tensor possessing the frequency and spatial dispersions. In the range of its applicability, i.e., for sufficiently large wavelengths, such a procedure gives the predictions that are in good agreement with full-fledged numerical simulations and experiments. As far as the media constituted by straight or helical conducting wires obtained from each other by a parallel transport are concerned, such an approach was developed in the papers \cite{Belov2003,Belov2003hel,Magarill2003,Silveirinha2008,Nesterenko2008,Maslovski2009,Morgado2012,Morgado_2012,Tyukhtin2014,Morgado2016,Yakovlev2020}. However, the effective permittivity tensor obtained and used in those papers turns out to be translation invariant even in the case of the medium composed of the conducting spiral wires. It is clear that the Maxwell equations with such an effective permittivity do not reproduce the Bragg resonances and the chiral forbidden band of the electromagnetic waves in helical media. In the present paper, inspired by the effective field theory approach, we generalize the results of these papers and describe the electromagnetic properties of metamaterials consisting of conducting spiral wires taking exactly into account the symmetry of the medium and carrying out the derivative expansion of the permittivity tensor and the expansion with respect to the powers of the photon momentum, $\spk$, near the plasmon resonance \cite{LandLifshECM}. In the degenerate case of straight conducting wires, the permittivity tensor obtained by us coincides with the effective permittivity tensor derived in the papers \cite{Belov2003,Silveirinha2008}. Adjusting correspondingly the parameters of the effective model, the dispersion law of the electromagnetic waves in our model coincides with the dispersion law obtained in the semianalytical model proposed in \cite{Wu2010} and with the rigorous numerical simulations presented in \cite{LiWongChan2015}. Moreover, the effective field theory approach predicts the existence and the form of the corrections to the effective permittivity tensor that are small for a rarefied wired medium but appear to be relevant for a close-packed arrangement of wires.

As a result of the strong spatial dispersion, the additional degree of freedom -- the plasmon field -- appears in the effective model. By introducing this field into the theory, we get rid of nonlocality of the permittivity tensor and obtain a simple local in space effective model. This model is exactly solvable in the paraxial limit, i.e., for vanishing transverse component of momentum of the fields in the medium. Furthermore, the constructed model admits an exact solution in the shortwave limit when the propagation velocity of plasmons is equal to the velocity of light. This exact solution is valid in the nonparaxial regime as well. Having found the complete set of solutions in the helical wired medium, we investigate the problem of scattering of plane electromagnetic waves by a slab made of such a medium. Then the problem of imposing the additional boundary conditions arises \cite{Pekar1958,Pekar21958,Agranovich1966,Maradudin1973,Agarwal1974,ArgMill82,Chen1993}. These additionally conditions are needed for the unique solution of the scattering problem. In the framework of the effective model we consider, this problem is naturally solved: one ought to require that the plasmon field is equal to zero out of the wired medium. This is the well-known additional boundary condition by Pekar \cite{Pekar1958,Pekar21958}. We will show that this boundary condition is equivalent to the additional boundary condition employed in \cite{Silverinha2006} in describing scattering of electromagnetic waves by a slab made of parallel straight conducting wires.

Since the effective model is exactly solvable, the dispersion law of the electromagnetic waves in helical wired medium can be analyzed for the different values of parameters of this medium. As a result, we describe in detail the band structure, the polarization properties, and the energy density flux for the different modes of the electromagnetic field. These properties allow one to provide a clear physical interpretation to the data of scattering of plane electromagnetic waves by a slab made of the helical wired medium. Notice that for the helix angle $\alpha\approx\pi/2$ the structure we investigate resembles a cholesteric, which is probably the most studied natural material with helical symmetry \cite{Belyakov1982,deGennes1993,Yang2006,Barboza2015,Belyakov2019,Vetrov2020,Li2021}. The major difference of the helical wired medium from the cholesterics is that the permittivity tensor of the former possesses a strong spatial dispersion. This leads, in particular, to the presence of the total forbidden band in the spectrum of photons for sufficiently small energies. The cholesterics do not have such a property.

The paper is organized as follows. In Sec. \ref{GenForm}, we start with the symmetry analysis of the dielectric permittivity tensor for helical media with spatial dispersion. We derive here the general expression for such a tensor. Then we particularize this general expression to the case of the medium made of conducting spirals, the spirals being obtained from each other by a parallel transport. Performing the derivative expansion and the expansion with respect to the photon momentum and keeping the helical symmetry intact, we derive the approximate expression for the permittivity tensor of helical wired medium. Introducing the additional plasmon field, we eventually formulate the effective model for electromagnetic fields in such a medium. In Sec. \ref{Normal}, we consider the exactly solvable paraxial case where the momentum component perpendicular to the helix axis is zero. We investigate the behavior of the dispersion law and its asymptotics, the polarization properties of the field modes in the medium, the structure of the forbidden bands, and the integral curves of the Poynting vector. In Sec. \ref{Shrt_Wave_Appr}, we conduct a similar analysis in the shortwave approximation abandoning the assumption of paraxial propagation of the electromagnetic fields in the medium. In Conclusion, we summarize the results. Throughout the paper we use the system of units such that $c=\hbar=1$.

\section{Effective model}\label{GenForm}

The Maxwell equations in a dispersive medium take the form
\begin{equation}\label{MaxwellEq}
	(\text{rot}^2_{ij} -k_0^2\hat{\e}_{ij})A_j=0,\qquad \hat{k}_i(\hat{\e}_{ij}A_j)=0,
\end{equation}
where $\hat{k}_i:=-i\partial_i$. The second equation is the Coulomb gauge. It follows from the first equation for $k_0\neq0$. Henceforth, we suppose that $k_0\neq0$ and, consequently, one may solve only the first equation.

We will study the propagation of electromagnetic waves in the helical medium possessing the frequency and spatial dispersions. In that case, $\hat{\e}_{ij}$ depends on $k_0$ and is a nonlocal operator with respect to the spatial variables. Denote the kernel of this operator as $K_{ij}(k_0;\spx,\spy)$. It is useful to describe such an operator by its Weyl symbol (see, e.g., \cite{Berezin1980}), $\e_{ij}(k_0;\spx,\spk)$, which is in one-to-one correspondence with the kernel of the operator
\begin{equation}\label{Weyl_symb_defn}
    \e_{ij}(k_0;\spx,\spk)=\int d\spy e^{i\spk\spy} K_{ij}\big(k_0;\spx-\frac{\spy}{2},\spx+\frac{\spy}{2}\big),\qquad K_{ij}(k_0;\spx,\spy)=\int\frac{d\spk}{(2\pi)^3} e^{i\spk(\spx-\spy)} \e_{ij}\big(k_0;\frac{\spx+\spy}{2},\spk\big).
\end{equation}
In the case when the spatial dispersion is absent the complete description of the helically symmetric permittivity tensor is presented in \cite{KK2022}. For the reader convenience, we recall here the notation introduced in that paper.

The operator of the total angular momentum is written as
\begin{equation}\label{ang_mom}
\begin{split}
    \hat{J}_{l ij}&:=\hat{L}_{l ij}+S_{l ij},\\
    \hat{L}_{l ij}&:=\e_{lmn}x_m \hat{k}_n\de_{ij},\qquad S_{l ij}:=-i\e_{l ij},
\end{split}
\end{equation}
where the index $l$ numerates the components of the angular momentum operator, $\hat{L}_{lij}$ is the operator of the orbital angular momentum, and $S_{l ij}$ is the spin operator of a photon. The operator of rotation by an angle of $\psi$ around the $z$ axis takes the form
\begin{equation}\label{R_psi}
    \hat{R}_\psi=\hat{R}^L_\psi R^S_\psi=e^{i\psi \hat{J}_3}=e^{i\psi \hat{L}_3}e^{i\psi S_3}.
\end{equation}
Let $\{\spe_1,\spe_2,\spe_3\}$ be a high-handed orthonormal triple. Then
\begin{equation}\label{epm_e3}
    S_3 \spe_\pm=\pm\spe_\pm,\qquad S_3\spe_3=0,
\end{equation}
where
\begin{equation}
    \spe_\pm:=\spe_1\pm i\spe_2.
\end{equation}
Any vector can be decomposed in terms of the eigenvectors of the operator $S_3$ as
\begin{equation}\label{pm_basis}
    \spx=\frac12(x_-\spe_++x_+\spe_-)+x_3\spe_3.
\end{equation}
Denote as $\hat{T}_{a}$ the translation operator along the $z$ axis: $z\rightarrow z+a$. Then the dielectric permittivity tensor, $\hat{\e}_{ij}$, enjoys the helical symmetry provided that
\begin{equation}\label{hel_permit}
    \hat{R}_\psi \hat{T}_{\psi/q} \hat{\e} (\hat{R}_\psi \hat{T}_{\psi/q})^{-1}=\hat{\e},\quad\forall\psi\in \R,
\end{equation}
where the tensor indices of $\hat{\e}$ are not shown explicitly and $2\pi/|q|$ defines the helix pitch.

As follows from the relation between the operator kernel and its Weyl symbol \eqref{Weyl_symb_defn}, the Weyl symbol of the helically symmetric permittivity tensor satisfies
\begin{equation}\label{hel_permit_symb}
    \hat{R}_\psi \hat{T}_{\psi/q} \e (\hat{R}_\psi \hat{T}_{\psi/q})^{-1}=\e,\quad\forall\psi\in \R,
\end{equation}
where the rotation operator, $\hat{R}_\psi$, acts on the variables $\spk$ in the same way as on the variables $\spx$, i.e., the operator of the orbital momentum in expression \eqref{R_psi} should be replaced by
\begin{equation}
    \hat{L}_{l ij}=\big(\e_{lmn}x_m \hat{k}_n -i\e_{lmn}k_m \frac{\partial}{\partial k_n}\big)\de_{ij}.
\end{equation}
The translation operator, $\hat{T}_{a}$, does not act on the variables $\spk$. As a result, we can just apply the analysis carried out in \cite{KK2022} and write the general expression for the Weyl symbol obeying \eqref{hel_permit_symb},
\begin{equation}\label{eps_hel_gen}
    \e=\sum_{s=-2}^2\sum_{l,l'=-\infty}^\infty e^{i(l\phi+l'\vf)}e^{-iq(l+l'+s)z}\e^{(s)}_{ll'},
\end{equation}
where $\phi:=\arg x_+$, $\vf:=\arg k_+$, and $\e^{(s)}_{ll'}$ are irreducible tensor of the spin $s$. They read
\begin{equation}\label{eps_hel_s}
\begin{split}
    \e^{(\pm2)}_{ll'}&=A^\pm_{ll'}\spe_\pm\otimes\spe_\pm,\\
    \e^{(\pm1)}_{ll'}&=\al^\pm_{ll'}\spe_\pm\otimes\spe_3+\be^\pm_{ll'}\spe_3\otimes\spe_\pm,\\
    \e^{(0)}_{ll'}&=\e_{ll'}\spe_+\spe_- +y_{ll'}\spe_+\wedge\spe_- +\e_{\perp ll'}\spe_3\otimes\spe_3,
\end{split}
\end{equation}
where
\begin{equation}
    \mathbf{a}\mathbf{b}:=\frac12(\mathbf{a}\otimes\mathbf{b}+\mathbf{b}\otimes\mathbf{a}),\qquad
    \mathbf{a}\wedge\mathbf{b}:=\frac12(\mathbf{a}\otimes\mathbf{b}-\mathbf{b}\otimes\mathbf{a}),
\end{equation}
and the coefficients standing at the tensors are some functions of $x_\perp$, $k_\perp$, $k_3$, and $k_0$.

The kernel of the dielectric permittivity operator possesses certain symmetries \cite{LandLifshECM}:
\begin{equation}
\begin{split}
    i)\;K^*_{ij}(k_0;\spx,\spy)&=K_{ij}(-k_0;\spx,\spy)\;\text{-- reality};\\
    ii)\;K_{ij}(k_0;\spx,\spy)&=\tilde{K}_{ji}(k_0;\spy,\spx)\;\text{-- symmetry of the kinetic coefficients};\\
    iii)\;K^*_{ij}(k_0;\spx,\spy)&=K_{ji}(k_0;\spy,\spx)\;\text{-- transparency of the medium},
\end{split}
\end{equation}
where the tilde means that one has to reverse the sign of the magnetic field strength. Formulas \eqref{Weyl_symb_defn} imply that these symmetry relations are equivalent to
\begin{equation}
\begin{split}
    i)\;\e^*_{ij}(k_0;\spx,\spk)&=\e_{ij}(-k_0;\spx,-\spk);\\
    ii)\;\e_{ij}(k_0;\spx,\spk)&=\tilde{\e}_{ji}(k_0;\spx,-\spk);\\
    iii)\;\e^*_{ij}(k_0;\spx,\spk)&=\e_{ji}(k_0;\spx,\spk),
\end{split}
\end{equation}
in terms of the symbol of the permittivity operator. These relations result in the constraints on the coefficients in \eqref{eps_hel_s}:
\begin{equation}
\begin{aligned}
    i)\;[A^+_{ll'}(k_0)]^*&=A^-_{-l,-l'}(-k_0),&\qquad [\al^+_{ll'}(k_0)]^*&=\al^-_{-l,-l'}(-k_0),& \qquad[\be^+_{ll'}(k_0)]^*&=\be^-_{-l,-l'}(-k_0),\\
    \e^*_{ll'}(k_0)&=\e_{-l,-l'}(-k_0),& \qquad y^*_{ll'}(k_0)&=-y_{-l,-l'}(-k_0),&\qquad \e^*_{\perp ll'}(k_0)&=\e_{\perp -l,-l'}(-k_0);\\
    ii)\; \tilde{A}^\pm_{ll'}(-k_3)&=(-1)^{l'}A^\pm_{ll'}(k_3),&\qquad \tilde{\al}^\pm_{ll'}(-k_3)&=(-1)^{l'}\be^\pm_{ll'}(k_3),& \qquad \tilde{\be}^\pm_{ll'}(-k_3)&=(-1)^{l'}\al^\pm_{ll}(k_3),\\
    \tilde{\e}_{ll'}(-k_3)&=(-1)^{l'}\e_{ll'}(k_3),& \qquad \tilde{y}_{ll'}(-k_3)&=-(-1)^{l'}y_{ll'}(k_3),&\qquad \tilde{\e}_{\perp ll'}(-k_3)&=(-1)^{l'}\e_{\perp ll'}(k_3);\\
    iii)\quad\;\;[A^+_{ll'}]^*&=A^-_{-l,-l'},&\qquad [\al^+_{ll'}]^*&=\be^-_{-l,-l'},& \qquad[\be^+_{ll'}]^*&=\al^-_{-l,-l'},\\
    \e^*_{ll'}&=\e_{-l,-l'},& \qquad y^*_{ll'}&=y_{-l,-l'},&\qquad \e^*_{\perp ll'}&=\e_{\perp -l,-l'}.
\end{aligned}
\end{equation}
If the symbol of the permittivity operator is smooth, then these coefficients must have the form
\begin{equation}
    c_{ll'}(x_\perp,k_\perp)=x_\perp^{|l|}k_\perp^{|l'|}f_{ll'}(x_\perp^2,k_\perp^2),
\end{equation}
where $f_{ll'}$ are some smooth functions. Of course, they are different for the different coefficients in the irreducible tensor. If the helical medium is invariant under translations in the $(x,y)$ plane, then one should put $l=0$ in expansion \eqref{eps_hel_gen} and take the coefficients in formulas \eqref{eps_hel_s} to be independent of $x_\perp$.

Let us consider the particular case of the helical medium comprised by thin conducting spiral wires immersed into the dielectric medium with permittivity $\e_h(k_0)$. The helix axis is directed along the $z$ axis and the spiral wires are obtained from each other by a parallel transport in the $(x,y)$ plane. We assume that
\begin{equation}\label{long_wave}
    \e^{1/2}_hk_0 a\ll1,
\end{equation}
where $a$ is the typical distance between the wires in the $(x,y)$ plane. The periodicity of positions of the wires in the $(x,y)$ plane is not assumed. Furthermore, it is supposed that the wires are uniformly distributed in the $(x,y)$ plane on the scales much larger than $a$. Moreover, we assume that the wires have a circular section and are sufficiently thin, $r_w\ll a$, where $r_w$ is the wire radius. The unit tangent vector to the wires is written as
\begin{equation}\label{xi_tau}
	\boldsymbol{\tau} (z)= \cos\alpha \spe_3+\sin\alpha\mathbf{d}(z),\qquad
    \mathbf{d}(z)=(\cos(qz),\sin(qz),0)^T,\quad \alpha\in[0,\pi/2].
\end{equation}
The vectors $\bs\tau$, $\mathbf{d}$, and $\spe_3$ are invariant under the helical transformations $\hat{R}_\psi \hat{T}_{\psi/q}$. Besides,
\begin{equation}
    \partial_z\bs\tau=q[\spe_3,\bs\tau].
\end{equation}
Assume additionally that
\begin{equation}\label{short_wave}
    \e^{1/2}_hk_0/q\gg1,
\end{equation}
i.e., we assume that the vector $\bs\tau$ changes slowly on the wavelength scale. We also suppose that the magnetic field produced by the medium or by the external sources does not enter into the permittivity tensor. In the latter case, it means that only the linear response on the external electromagnetic field is taken into account.

In the longwave limit \eqref{long_wave} under the assumptions above, we conclude that the effective permittivity tensor of the medium is translation invariant in the $(x,y)$ plane. Moreover, as long as it is assumed that the wires are thin and have a circular section, the effective permittivity tensor must be constructed in terms of the vectors $\bs\tau$, $\spk$, and the tensors $\de_{ij}$ and $\e_{ijk}$. In other words, locally there are only two distinguished vectors $\bs\tau$ and $\spk$ in the medium. The condition \eqref{short_wave} implies that the contributions of the vectors $\partial^n_z\bs\tau$ can be neglected. Usually the spatial dispersion is small and only near resonances or in conducting media can its influence be determinative (\cite{LandLifshECM}, Sec. 106). Therefore, we keep the dependence of the permittivity tensor on $\spk$ only in the terms that specify the resonances, i.e., in the denominators of the coefficients standing at the tensor structures in \eqref{eps_hel_s}. Demanding the fulfillment of the symmetry properties $(i)-(iii)$ and using the available set of invariant tensors, we can expand the denominators near the resonance in the leading order in $\spk$ as
\begin{equation}
    \e_hk_0^2-m^2-v^2(\bs\tau\spk)^2-b^2\spk^2,
\end{equation}
where the coefficient $\e_h$ at $k_0^2$ is included just for convenience. In virtue of the property $(ii)$, the tensors that can be employed in construction of $\e_{ij}$ should be symmetric, i.e., in our case these tensors are $\de_{ij}$ and $\tau_i\tau_j$. The contribution with $\de_{ij}$ describes the isotropic response of the medium on the external electromagnetic field. Inasmuch as the polarization current should be mainly directed along $\bs\tau$ for conducting wires, the coefficient at $\de_{ij}$ enjoys only a frequency dispersion and does not have resonances caused by conductivity of spiral wires. As as result, the symbol of the permittivity tensor operator becomes
\begin{equation}
	\e_{ij}(k_0; \spx,\spk)=\e \Big[ \de_{ij}-  \frac{ \omega_p^2\tau_i\tau_j}{
    \e_hk_0^2-m^2-v^2(\bs\tau\spk)^2-b^2\spk^2}\Big],
\end{equation}
where $\omega_p$ is some parameter that we call the plasma frequency. The poles of the permittivity tensor are related to the presence of the additional degrees of freedom of the electromagnetic field in the medium at issue. These are the plasmon-polaritons. The zeros of the denominator determine the dispersion law of the plasmons. It is seen from the resulting dispersion law that the coefficient $b$ is responsible for the isotropic propagation of plasmons in all directions including the directions perpendicular to the conducting wires. Since we assume that the wires are thin and the distance between them is large as compared with the wire radius, this contribution to the dispersion law can be neglected, i.e., $b\approx0$. By assumption, the conducting wires occupy a small volume in the dielectric medium. Hence, $\e\approx\e_h$. The effective mass of plasmons, $m$, must be close to zero, because the resonance resulting from conductivity of wires, i.e., due to creation of plasmons inducing the macroscopic polarization current along $\bs\tau$, appears at an arbitrarily small frequency of the external electromagnetic field. The plasma frequency $\omega_p$ cannot be found from the general considerations. However, it should be of order $1/a$ on dimensional grounds.

Therefore, up to the terms of higher order in derivatives of the vector $\bs\tau$, we have eventually the kernel of the effective permittivity tensor
\begin{equation}\label{diel_permit}
	K_{ij}(k_0; \spx,\spx')=\e_h \Big[ \de_{ij}- \tau_i(z) \frac{ \omega_p^2}{\omega_0^2
    -v^2(\boldsymbol{\tau}(z)\hat{\spk})^2}\tau_j(z')\Big]\de(\spx-\spx'),
\end{equation}
where $v$ is interpreted as the velocity of propagation of plasmons along the conducting wires and, for brevity, the notation has been introduced $\omega_0:=\e_h^{1/2} k_0$. Such an expression for the permittivity tensor of the helical wired medium can also be obtained using the results of the papers \cite{Belov2003,Silveirinha2008} where the effective permittivity tensor for an array of parallel straight wires was deduced. To this end, supposing that condition \eqref{short_wave} is satisfied, we imagine that one cuts the array of conducting wires into the layers perpendicular the $z$ axis of the width much less than $1/|q|$, for the conducting wires can be approximately considered as straight, and much larger than the wavelength of the electromagnetic field, for the edge effects can be neglected. The kernel of the permittivity tensor operator for every such layer takes the form \eqref{diel_permit} with the constant vector $\bs\tau$ directed along the straight conducting wires \cite{Belov2003,Silveirinha2008}. In these papers, the explicit expressions for the plasma frequency, $\omega_p$, of the array of parallel straight wires are given. This frequency is indeed of order  $1/a$.

We can get rid of nonlocality in the Maxwell equations \eqref{MaxwellEq} with the permittivity tensor \eqref{diel_permit} by introducing the additional scalar field obeying certain boundary conditions. It is easy to see that the initial equations \eqref{MaxwellEq}, \eqref{diel_permit} are equivalent to the system
\begin{equation}\label{MaxwellEq0}
	\begin{split}
		(\omega_0^2 - v^2 (\boldsymbol{\tau}\hat{\spk})^2) \Psi + \omega_0 \omega_p (\boldsymbol{\tau}\mathbf{A})&=0,\\
		(\omega_0^2-\text{rot}^2)\mathbf{A}+\omega_0\omega_p\Psi\boldsymbol{\tau}&=0,
	\end{split}
\end{equation}
where $\Psi$ is the scalar field of plasmons existing only in the wired medium. The equations \eqref{MaxwellEq0} can be obtained by varying the action functional
\begin{equation}
    S[\mathbf{A},\mathbf{A}^*,\Psi,\Psi^*]=\int d\spx\big\{\mathbf{A}^*(\omega_0^2-\text{rot}^2)\mathbf{A} + \Psi^*(\omega_0^2 - v^2 (\boldsymbol{\tau}\hat{\spk})^2) \Psi +\omega_0\omega_p [\Psi^*(\bs\tau\mathbf{A}) +(\bs\tau\mathbf{A}^*)\Psi ]\big\}.
\end{equation}
The vector,
\begin{equation}
    \mathbf{j}_{pol}:=\omega_0\omega_p\Psi\boldsymbol{\tau},
\end{equation}
can be interpreted as the polarization current density of the wired medium.

In the present paper, we assume that the helical wired medium fills an infinite plate of the width $L$ normal to the $z$ axis. It is located at $z\in(0,L)$. The boundary conditions providing self-adjointness of the operator acting on the fields $(\mathbf{A},\Psi)$ in the system of equations \eqref{MaxwellEq0} read
\begin{equation}\label{PsiBoundCond}
	[\mathbf{A}_\perp]_{z=0}=[\mathbf{A}_\perp]_{z=L}=0,\qquad[\rot\mathbf{A}_\perp]_{z=0}=[\rot\mathbf{A}_\perp]_{z=L}=0
    \qquad\Psi(0)=\Psi(L)=0,
\end{equation}
where $\mathbf{A}_\perp$ and $\rot\mathbf{A}_\perp$ denote the $(x,y)$ components of the corresponding vectors and the square brackets mean a discontinuity jump of the corresponding quantity on the surface indicated. The first two conditions are the standard boundary conditions imposed on the vector potential on the interface between two media (see, e.g., \cite{RyazanovB,LandLifshECM}). The last boundary condition is the standard boundary condition for the plasmon field \cite{Pekar1958,Pekar21958}. Notice that the other boundary conditions for electromagnetic fields in wired media were also proposed \cite{Silverinha2006}. In Appendix \ref{Bound_Cond_Ap}, we prove that the additional boundary conditions presented in \cite{Silverinha2006} are in fact equivalent the boundary conditions \eqref{PsiBoundCond}. Further, we suppose that the slab made of the wired medium is placed in the isotropic dielectric with permittivity $\e_0(k_0)$.

As long as the system we study is translation invariant in the $(x,y)$ plane, we seek for a solution of \eqref{MaxwellEq0} in the form
\begin{equation}\label{Anzatz1}
    \mathbf{A}(\mathbf{x})=e^{i\mathbf{k}_\perp \mathbf{x}_\perp}\mathbf{A}(z),\qquad \Psi(\mathbf{x})=e^{i\mathbf{k}_\perp \mathbf{x}_\perp}\Psi(z).
\end{equation}
Introduce the notation for the components of the electromagnetic field potential in the basis \eqref{pm_basis}:
\begin{equation}\label{Anzatz2}
    A_\pm (z)=a_\pm (z)e^{\pm i \vf},\qquad A_3(z)=a_3(z),
\end{equation}
where $\vf = \arg k_+$. The component $A_3(z)$ can be expressed from the second equation in \eqref{MaxwellEq0} as
\begin{equation}\label{A_3}
    A_3(z)=a_3(z)=- \cos \alpha\frac{\omega_0\omega_p}{\bar{k}_3^2}\Psi(z) -\frac{k_\perp}{2\bar{k}_3^2}\hat{k}_3(a_+(z)+a_-(z)),
\end{equation}
where $\bar{k}_3:=(\omega_0^2-k_\perp^2)^{1/2}$. Then the Maxwell equations are reduced to the system of three linear ordinary differential equations of the second order
\begin{equation}\label{MaxwellEq1}
\begin{split}
    \omega_0^2 a_+ -\frac{k_\perp^2}{2}(a_+-a_-)-\frac{k_\perp^2}{2\bar{k}^2_3}\hat{k}_3^2(a_++a_-)-\hat{k}_3^2a_+ +\omega_p \omega_0\big[\sin\al e^{i\theta} -\frac{k_\perp\cos\al}{\bar{k}_3^2}\hat{k}_3 \big]\Psi&=0,\\
    \omega_0^2 a_- +\frac{k_\perp^2}{2}(a_+-a_-)-\frac{k_\perp^2}{2\bar{k}^2_3}\hat{k}_3^2(a_++a_-)-\hat{k}_3^2a_- +\omega_p \omega_0\big[\sin\al e^{-i\theta} -\frac{k_\perp\cos\al}{\bar{k}_3^2}\hat{k}_3 \big]\Psi&=0,\\
    \big[\omega_0^2 -v^2(\cos\al\hat{k}_3 +k_\perp\sin\al\cos\theta)^2 -\frac{\cos^2\al \omega_p^2\omega_0^2}{\bar{k}_3^2} \big]\Psi +\frac{\omega_p\omega_0}{2}\big[\sin\al (e^{i\theta} a_-+e^{-i\theta}a_+) -\frac{k_\perp\cos\al}{\bar{k}_3^2}\hat{k}_3(a_++a_-) \big]&=0,
\end{split}
\end{equation}
where $\theta:=qz-\vf$. On changing the normalization of the field $\Psi$ and introducing $\tilde{\Psi}:=\sqrt{2}\Psi$, the system \eqref{MaxwellEq1} is bring into the explicitly self-adjoint form
\begin{equation}\label{MaxwellEq2}
    \big[\hat{k}_3 M_0\hat{k}_3+\frac12(\hat{k}_3 M_1+M_1\hat{k}_3) +M_2\big]W=0,
\end{equation}
where $W^T=[a_+,a_-,\tilde{\Psi}]$ and $M_0^\dag=M_0$, $M_1^\dag=M_1$, $M_2^\dag=M_2$. Besides,
\begin{equation}
\begin{split}
    M_0&=-
    \left[
      \begin{array}{ccc}
        1+\frac{k_\perp^2}{2\bar{k}_3^2} & \frac{k_\perp^2}{2\bar{k}_3^2} & 0 \\
        \frac{k_\perp^2}{2\bar{k}_3^2} & 1+\frac{k_\perp^2}{2\bar{k}_3^2} & 0 \\
        0 & 0 & v^2\cos^2\al \\
      \end{array}
    \right],\\
    M_1&=-
    \left[
      \begin{array}{ccc}
        0 & 0 & \frac{\omega_p\omega_0k_\perp\cos\al}{\sqrt{2}\bar{k}^2_3} \\
        0 & 0 & \frac{\omega_p\omega_0k_\perp\cos\al}{\sqrt{2}\bar{k}^2_3} \\
        \frac{\omega_p\omega_0k_\perp\cos\al}{\sqrt{2}\bar{k}^2_3} & \frac{\omega_p\omega_0k_\perp\cos\al}{\sqrt{2}\bar{k}^2_3} & v^2k_\perp\sin(2\al)\cos\theta \\
      \end{array}
    \right],\\
    M_2&=
    \left[
      \begin{array}{ccc}
        \omega_0^2-\frac{k_\perp^2}{2} & \frac{k_\perp^2}{2} & \frac{\omega_p}{\sqrt{2}}\omega_0\sin\al e^{i\theta} \\
        \frac{k_\perp^2}{2} & \omega_0^2-\frac{k_\perp^2}{2} & \frac{\omega_p}{\sqrt{2}}\omega_0\sin\al e^{-i\theta} \\
        \frac{\omega_p}{\sqrt{2}}\omega_0\sin\al e^{-i\theta} & \frac{\omega_p}{\sqrt{2}}\omega_0\sin\al e^{i\theta} & \omega_0^2 -\frac{\omega_p^2\omega_0^2}{\bar{k}_3^2}\cos^2\al -v^2k_\perp^2\sin^2\al\cos^2\theta \\
      \end{array}
    \right].
\end{split}
\end{equation}
The necessity of changing the normalization of the field $\Psi$ is a consequence of nonunitarity of the transformation from the field components $A_{1,2}$ to $A_\pm$.

In the general case, it seems impossible to find the exact solution to system \eqref{MaxwellEq2} with boundary conditions \eqref{PsiBoundCond} is a closed form. Nevertheless, these equations are exactly solvable in the paraxial limit when the perpendicular component of momentum of the electromagnetic fields vanishes (see Sec. \ref{Normal}). Furthermore, one can obtain the solution to the system \eqref{MaxwellEq0} in the shortwave approximation in the case $v=1$ (see Sec. \ref{Shrt_Wave_Appr}).

\section{Paraxial approximation}\label{Normal}
\subsection{Arbitrary helix angle $\al$}

Consider the propagation of an electromagnetic wave in a helical wired medium along the $z$ axis. For $k_\perp=0$, the system of ordinary differential equations \eqref{MaxwellEq2} is simplified to
\begin{equation}\label{MaxwellEqParax}
\begin{split}
    \omega_0^2 a_+ -\hat{k}_3^2a_+ +\frac{\omega_p \omega_0}{\sqrt{2}}\sin\al e^{i\theta} \tilde{\Psi}&=0,\\
    \omega_0^2 a_- -\hat{k}_3^2a_- +\frac{\omega_p \omega_0}{\sqrt{2}}\sin\al e^{-i\theta}\tilde{\Psi}&=0,\\
    \big[\omega_0^2 -(\omega_p^2+v^2\hat{k}^2_3)\cos^2\al  \big]\tilde{\Psi} +\frac{\omega_p\omega_0}{\sqrt{2}}\sin\al (e^{i\theta} a_-+e^{-i\theta}a_+)&=0.
\end{split}
\end{equation}
This system of equations has six linearly independent solutions. Due to periodicity of the coefficients of the system, it is useful to seek for a solution in the form
\begin{equation}\label{ParaxAnzatz}
    a_\pm = a_\pm(k_3)e^{i (k_3 \pm q)\theta/q}, \qquad \tilde{\Psi}=\tilde{\Psi}(k_3)e^{i k_3\theta/q},
\end{equation}
where $a_\pm(k_3)$ and $\tilde{\Psi}(k_3)$ are the coefficients independent of $z$, and $k_3$ is the physical momentum of the mode. The physical momentum of the mode is related to the quasimomentum $\kappa_3$ as $k_3=\kappa_3+q n$, $n\in \mathbb{Z}$. Substituting \eqref{ParaxAnzatz} into \eqref{MaxwellEqParax}, we arrive at the system of linear homogeneous algebraic equations
\begin{equation}\label{ParaxFourierCoef}
\begin{bmatrix}
    \omega_0^2 - (k_3+q)^2 & 0 & \frac{\omega_p \omega_0}{\sqrt{2}}\sin\al \\
    0& \omega_0^2 - (k_3-q)^2 & \frac{\omega_p \omega_0}{\sqrt{2}}\sin\al\\
    \frac{\omega_p \omega_0}{\sqrt{2}}\sin\al &  \frac{\omega_p \omega_0}{\sqrt{2}}\sin\al & \omega_0^2- (\omega_p^2 + v^2 k_3^2)\cos^2\alpha
\end{bmatrix}
\begin{bmatrix}
    a_+(k_3)\\
    a_-(k_3)\\
    \tilde{\Psi}(k_3)\\
\end{bmatrix}=0.
\end{equation}
The system obtained possesses nontrivial solutions when its determinant is zero. From this condition we obtain the dispersion law
\begin{equation}\label{ParaxDisp}
\begin{split}
    \omega_0^2 \omega_p^2 \sin^2\alpha(\omega_0^2-q^2-k_3^2)-\big(\omega_0^2-(k_3+q)^2\big)
    \big(\omega_0^2-(k_3-q)^2\big) \big(\omega_0^2- (\omega_p^2+k_3^2 v^2)\cos^2\alpha\big)=0.
\end{split}
\end{equation}
This dispersion law implicitly determines the dependence $\omega_0(k_3)$ or $k_3(\omega_0)$. The dispersion relation \eqref{ParaxDisp} is a third degree polynomial equation with respect to both $\omega_0^2$ and $k_3^2$. Therefore, it is possible to write out explicitly the dispersion law but the expressions turn out to be rather cumbersome. Below we will analyze equation \eqref{ParaxDisp} and give the explicit expressions for the dispersion law in some particular cases making certain approximations.

\begin{figure}[tp]
	\centering
	\includegraphics*[width=0.26\linewidth]{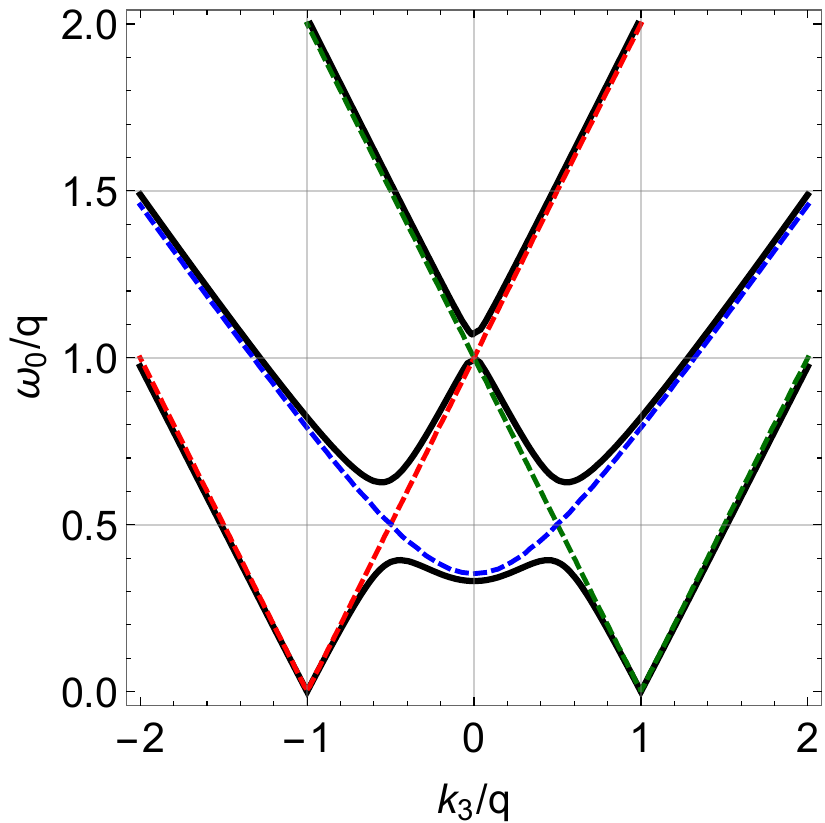}\;
	\includegraphics*[width=0.26\linewidth]{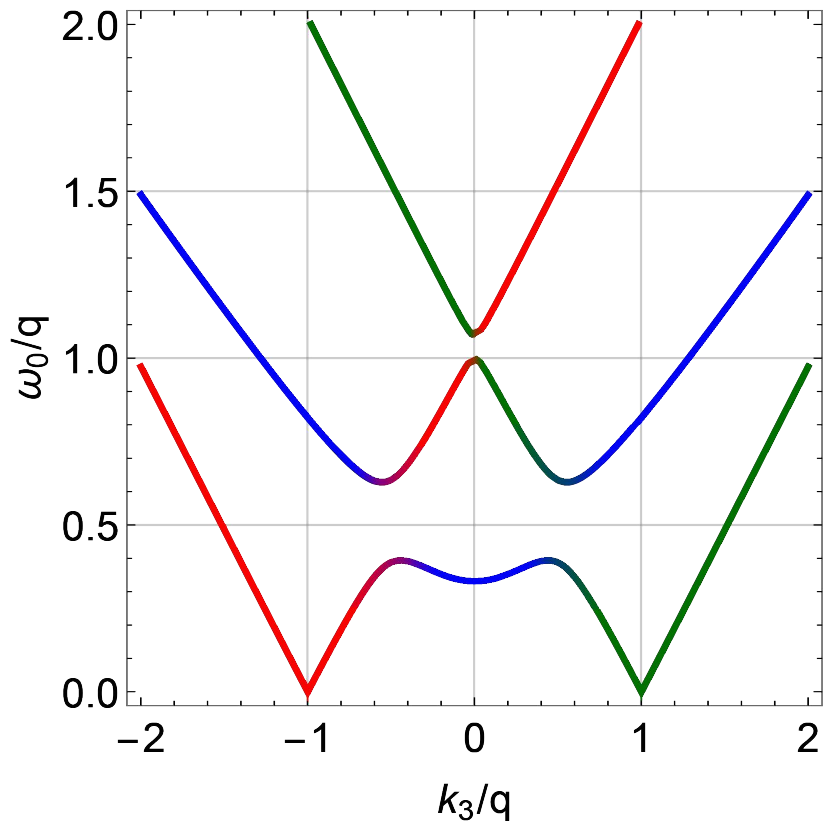}\;
	\includegraphics*[width=0.405\linewidth]{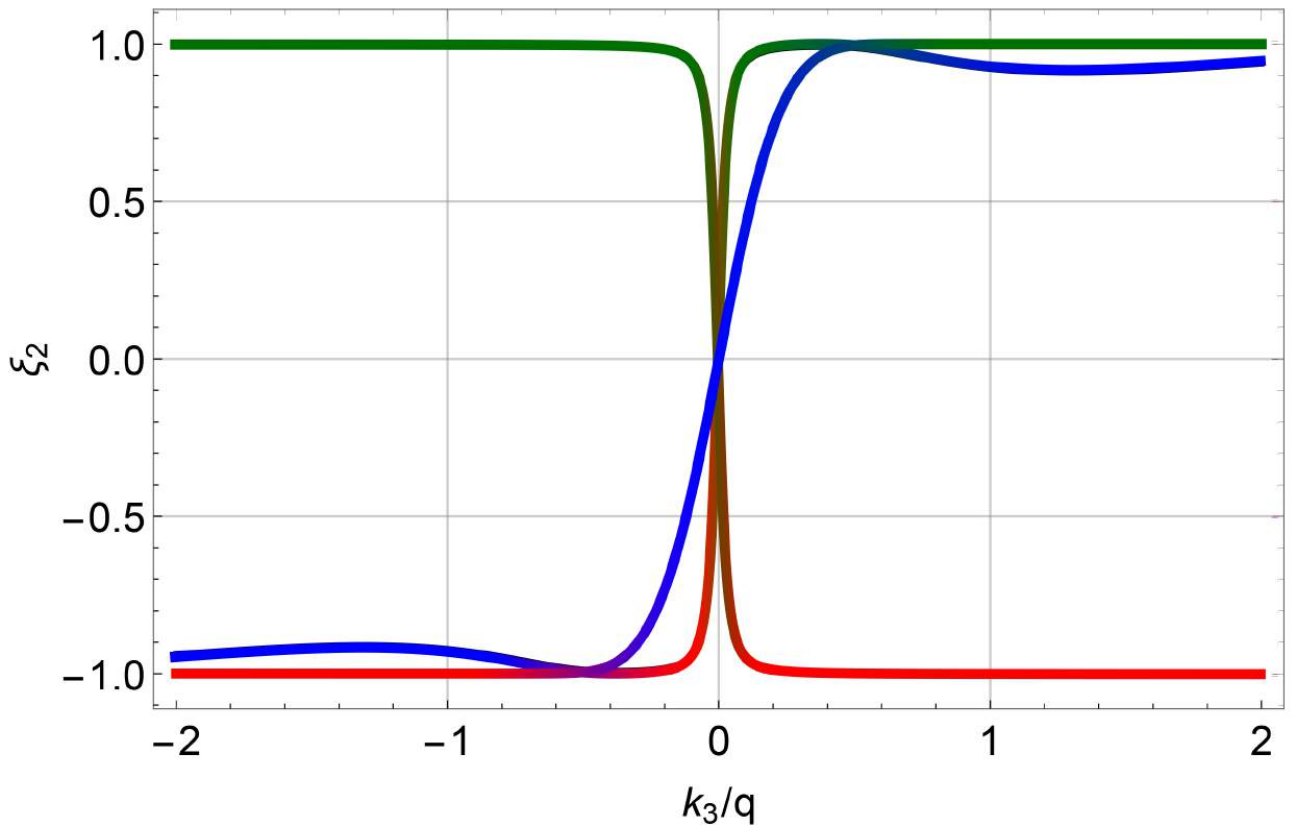}
	\caption{{\footnotesize The dispersion law and the Stokes parameter $\xi_2$ for the different modes of electromagnetic field in the helical wired medium \eqref{diel_permit} in the paraxial limit. The parameters are taken as follows: $\alpha = \pi/4$, $v=1$, $\omega_p=0.5$, $\e_h=1$, and $q=1$. Left panel: The exact branches of the dispersion law are plotted by the solid lines, whereas their asymptotics \eqref{ParaxDispAsym} are depicted by the dotted ones. Middle panel: The exact branches of the dispersion law are plotted by different colors showing their origin under the continuous deformation of the asymptotic branches of the dispersion law \eqref{ParaxDispAsym}. Right panel: The Stokes parameter $\xi_2$ for the different branches of the dispersion law. The colors on the plot of $\xi_2$ agree with the colors on the plot of the exact branches of the dispersion law. The Stokes parameter $\xi_2$ does not depend on the point $z$ in the medium. The branches of the dispersion law close to the cholesteric ones (green and red) possess a circular polarization. The polariton branch (blue) also has a high degree of circular polarization away from the point $k_3=0$. For the energies close to $\omega_0=0.5|q|$ and $\omega_0=|q|$, there are the chiral forbidden bands.}}
	\label{DispAsymp}
\end{figure}

\begin{figure}[tp]
	\centering
	\includegraphics*[width=0.39\linewidth]{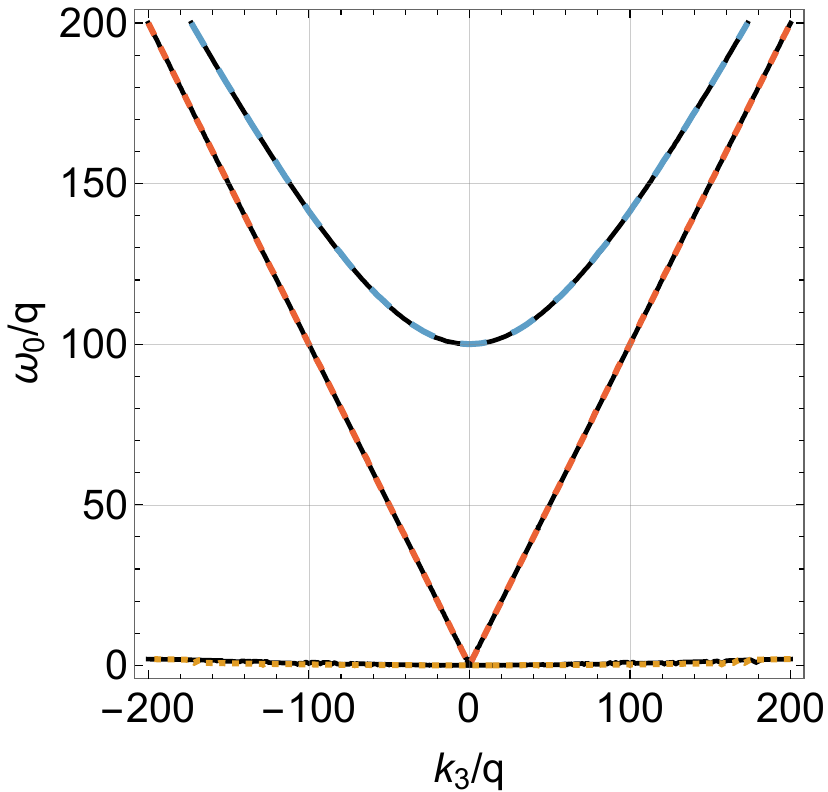}
	\includegraphics*[width=0.59\linewidth]{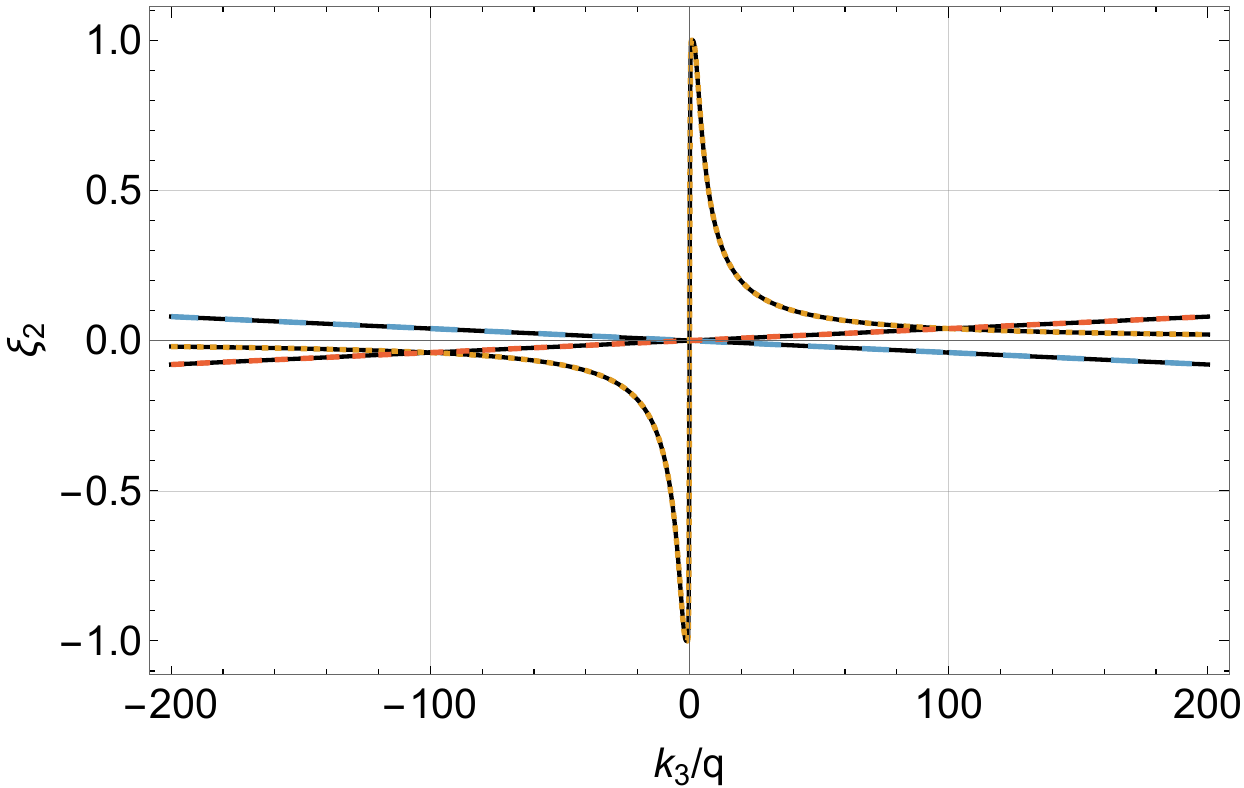}
	\caption{{\footnotesize  The dispersion law and the Stokes parameter $\xi_2$ for the different modes of electromagnetic field in the helical wired medium \eqref{diel_permit} in the shortwave paraxial limit. The parameters are taken as follows: $\pi/2-\alpha=0.01$, $v=1$, $\omega_p=100$, $\e_h=1$, and $q=1$. The exact branches of the dispersion law are plotted by the solid lines, whereas their asymptotics \eqref{wkb_disp} are depicted by the dotted ones. One can see good agreement between the approximate and exact expressions. For $|k_3|\sim\omega_p\gg|q|$, the modes of electromagnetic field are almost completely linearly polarized. Since $\al$ is close to $\pi/2$, one of the branches of the dispersion law is close to zero and the corresponding modes of electromagnetic field have a small group velocity along the $z$ axis (see \eqref{CholLikeDisp}).}}
	\label{DispWKB}
\end{figure}

The solution to system \eqref{ParaxFourierCoef} is given by
\begin{equation}\label{ParaxSol}
    a_\pm^{(n)} := a_\pm(k_3^{(n)})=-\frac{\omega_p \omega_0 \sin\al}{\sqrt{2}(\omega_0^2-(k_3^{(n)}\pm q)^2)},\qquad \tilde{\Psi} = 1,
\end{equation}
where $k_3^{(n)} \equiv k_3^{(n)}(\omega_0)$ is the solution to the dispersion equation \eqref{ParaxDisp} and the index $n=\overline{1,6}$ numerates these solutions. The different solutions of system \eqref{ParaxFourierCoef} are the different branches of a single multivalued analytic function of the complex variable $\omega_0$. Thus, we have found all the linearly independent solutions to the system of equations \eqref{MaxwellEqParax}. The expression for the electromagnetic potential is obtained by substituting \eqref{ParaxSol} into \eqref{Anzatz1}, \eqref{Anzatz2}. The result is
\begin{equation}\label{part_sol_n}
    \mathbf{A}^{(n)}(k_0,\spx)=-\frac{\omega_p}{\sqrt{2}\omega_0} e^{ik^{(n)}_3\theta/q}\Big[\cos\al\spe_3 +\frac{\sin\al}{2}\Big(\frac{\omega_0^2}{\omega_0^2-(k_3^{(n)}-q)^2} e^{-iqz}\spe_+ +\frac{\omega_0^2}{\omega_0^2-(k_3^{(n)}+q)^2} e^{iqz}\spe_- \Big) \Big].
\end{equation}
The general solution of the system of equations \eqref{MaxwellEq0} in the paraxial limit is a linear combination of the partial solutions \eqref{part_sol_n} with arbitrary constant coefficients.

Let us obtain the explicit expressions for the dispersion law in different asymptotic regimes. The analysis of the dispersion relation \eqref{ParaxDisp} reveals that if the following estimate is satisfied,
\begin{equation}\label{estim_asympt}
k_3^2+q^2 \gg \omega_p^2\sin^2\alpha,
\end{equation}
then the dispersion law for the different modes has the form
\begin{equation}\label{ParaxDispAsym}
\begin{split}
    \omega^2_0(k_3) &\approx \sigma^2_\pm,\qquad \sigma_\pm:= |k_3\mp q|,\\
    \omega^2_0(k_3) &\approx \sigma^2_p,\qquad \sigma_p:= \cos\alpha\sqrt{\omega_p^2+k_3^2v^2}.
\end{split}
\end{equation}
It is easy to see that these asymptotics are obtained by equating to zero the second term in \eqref{ParaxDisp}. They provide a good approximation for the dispersion law outside the neighborhoods of the intersection points of the branches $\sigma_n=\sigma_{n'}$, $n,n' \in \{\pm, p\}$ (see Fig. \ref{DispAsymp}), i.e., far from the regions of strong interaction of the dispersion law branches where they possess branch points as the functions of the complex variable $k_3$ or $\omega_0$.

The first term in \eqref{ParaxDisp} is responsible for mixing the branches of the dispersion law \eqref{ParaxDispAsym}. Due to the presence of branch points, there is an ambiguity in the choice of branches of the exact dispersion law. In this case, it is natural to consider the exact branches of the dispersion law as a continuous deformation of the asymptotic branches $\sigma_{\pm,p}$ (see Fig. \ref{DispAsymp}). Then the polarization properties of the corresponding modes of electromagnetic field do not differ considerably from the polarization properties of the modes taken on the asymptotic branches of $\sigma_{\pm,p}$ for a small deformation parameter, i.e., when the estimation \eqref{estim_asympt} is fulfilled.

\begin{figure}[tp]
	\centering
	\includegraphics*[width=0.37\linewidth]{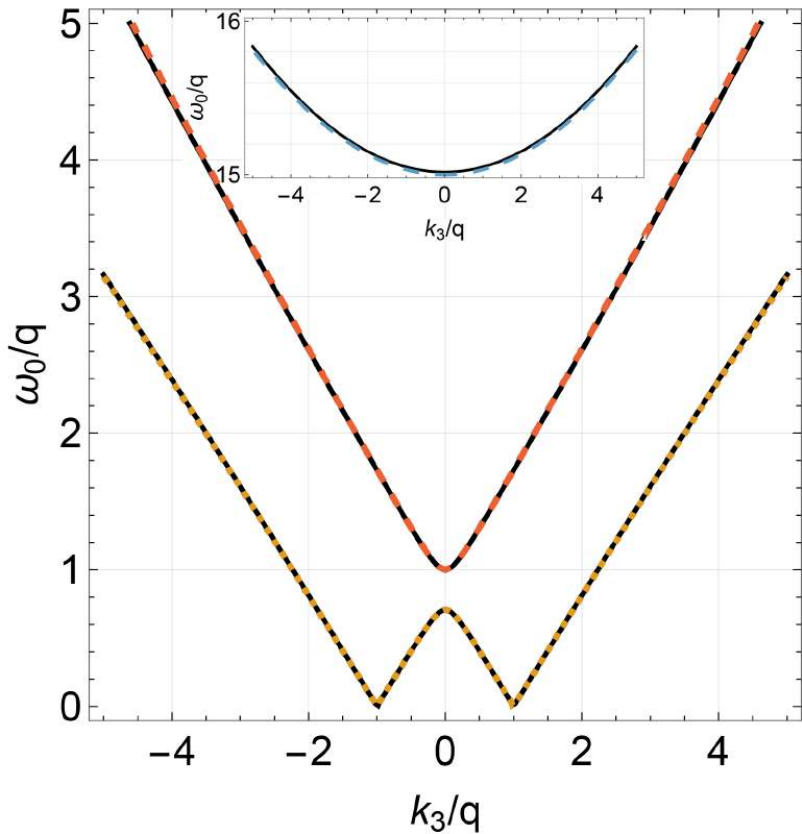}
	\includegraphics*[width=0.6\linewidth]{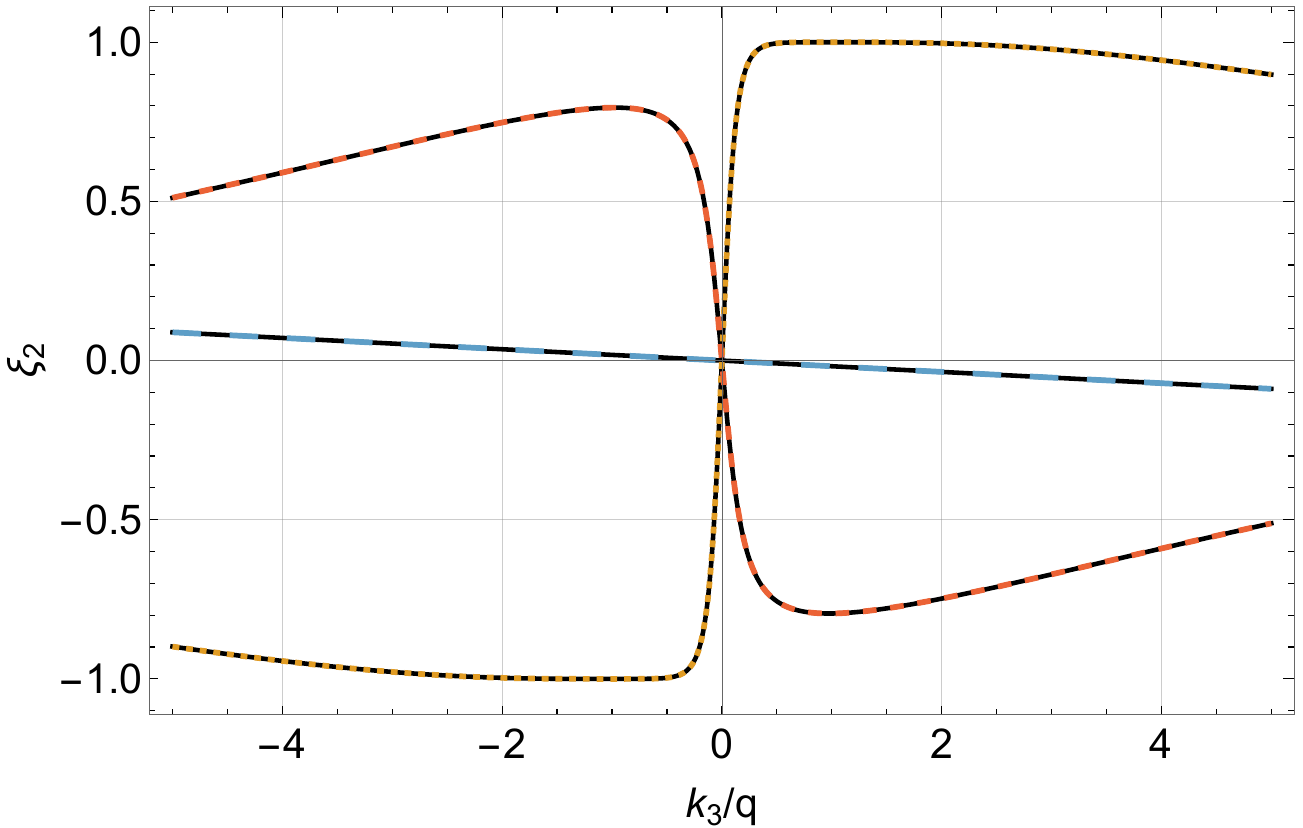}
	\caption{{\footnotesize
The dispersion law and the Stokes parameter $\xi_2$ for the different modes of electromagnetic field in the helical wired medium \eqref{diel_permit} in the paraxial limit. The parameters are taken as follows: $\alpha=\pi/4$, $v=1$, $\omega_p=15$, $\e_h=1$, and $q=1$. The solid lines indicate the exact dispersion law and the dotted lines show the approximate solution given by formula \eqref{disp_parax_lrg_omegap} and by the second formula in \eqref{wkb_disp_smlk3}. One can see good agreement between the approximate and exact expressions.}}
	\label{DispColored}
\end{figure}

The electromagnetic field modes corresponding to the branches of the dispersion law $\sigma_\pm$ have the same properties as the analogous branches of the dispersion law in cholesterics -- the natural helical media \cite{deGennes1993,Yang2006,Belyakov1982,Lakhtakia1995,Belyakov2019}. The branch $\sigma_p$ arises due to the presence of strong spatial dispersion. It is absent in the case of cholesterics. For definiteness, we refer to the branches of the dispersion law close to $\s_\pm$ as cholesteric and the branch of the dispersion law close to $\sigma_p$ as polariton. Notice that the three (without taking into account the change in the sign of $\omega_0$) different interacting branches of the dispersion law also arise in describing the propagation of sound waves in spiral anisotropic elastic media \cite{Oldano2000}. Two of these branches describe the propagation of transverse phonons whereas the third one describes the propagation of longitudinal phonons.

Let us consider the other asymptotic regimes when the dispersion law has a simple form. For large momenta and large plasma frequencies, i.e., under assumptions
\begin{equation}\label{wkb_regime}
    |k_3|\gg|q|,\qquad \omega_p \gg |q|,
\end{equation}
we have
\begin{equation}\label{wkb_disp}
\begin{split}
    \omega_0^2 &\approx \frac{1}{2}\Big[\omega_p^2+k_3^2(1+v^2\cos^2\alpha)\pm\sqrt{(\omega_p^2+k_3^2(1+v^2\cos^2\alpha))^2-4 k_3^2 (\omega_p^2+v^2 k_3^2)\cos^2\alpha}\Big],\\
    \omega_0^2&\approx k_3^2.
\end{split}
\end{equation}
In particular, for $v=1$, we obtain
\begin{equation}
    \omega_0^2\approx k_3^2\cos^2\alpha,\qquad \omega_0^2\approx \omega_p^2+k_3^2,\qquad \omega_0^2\approx k_3^2.
\end{equation}
The comparison of the exact dispersion law with the approximate one \eqref{wkb_disp} is shown in Fig. \ref{DispWKB}. If in addition to conditions \eqref{wkb_regime} we suppose that $|k_3|\gg\omega_p$, then
\begin{equation}
    \omega_0^2\approx v^2k_3^2\cos^2\al+\frac{(1-v^2)\omega_p^2\cos^2\al}{1-v^2\cos^2\al},\qquad \omega_0^2\approx k_3^2+\frac{\omega_p^2\sin^2\al}{1-v^2\cos^2\al},\qquad\omega_0^2\approx k_3^2.
\end{equation}
If $|k_3|\ll\omega_p$ and conditions \eqref{wkb_regime} are met, then
\begin{equation}\label{wkb_disp_smlk3}
\omega_0^2\approx k_3^2 \cos^2\al,\qquad \omega_0^2\approx\omega_p^2+(\sin^2\al+v^2\cos^2\al)k_3^2, \qquad\omega_0^2\approx k_3^2.
\end{equation}
In the next section, the parameter domain \eqref{wkb_regime} will be studied within the framework of shortwave approximation in the general case, i.e., for any $\spk_\perp$.

We also consider the regime which is the opposite to \eqref{estim_asympt}. Suppose that
\begin{equation}\label{large_omega_p}
    \omega_0^2\ll\omega_p^2\cos^2\al,\qquad v^2k_3^2\ll\omega_p^2.
\end{equation}
Then the dispersion equation \eqref{ParaxDisp} reduces to a quadratic equation for $\omega_0^2$ or $k_3^2$. The corresponding branches of the dispersion law become
\begin{equation}\label{disp_parax_lrg_omegap}
    \omega_0^2=\frac12\Big[(1+\cos^2\al)(q^2+k_3^2)\pm\sqrt{(q^2+k_3^2)^2\sin^4\al+16q^2k_3^2\cos^2\al}\Big].
\end{equation}
Notice that this approximate expression does not depend on $v$ and $\omega_p$. As for the remaining third branch, in this regime we can take the approximate expression given in the second formula in \eqref{wkb_disp_smlk3}. As seen in Fig. \eqref{DispColored}, the approximate expression \eqref{disp_parax_lrg_omegap} describes well the behavior of the dispersion law in the parameter domain \eqref{large_omega_p} including the vicinity of the branch points of the exact dispersion law and the chiral forbidden band (see Fig. \eqref{DispAsymp} and below).

Let us consider the polarization of electromagnetic waves corresponding to the different branches of the dispersion law. With that end in view, we introduce the Stokes parameters
\begin{equation}\label{Stokes_param}
    \xi_3+i\xi_1=\frac{2 A_+ A_-^*}{|A_+|^2+|A_-|^2},\qquad \xi_2=\frac{|A_-|^2-|A_+|^2}{|A_+|^2+|A_-|^2}.
\end{equation}
It is clear from the explicit expression for $A_\pm$ that
\begin{equation}\label{xi_plasmon}
    \xi_3+i\xi_1=re^{2iqz}\equiv\big(1-\frac{2a^2}{a^2+c^2}\big)e^{2iqz}, \qquad\xi_2=\frac{2ac}{a^2+c^2},
\end{equation}
where
\begin{equation}
    a:=2qk_3,\qquad c:=k_3^2+q^2-\omega_0^2.
\end{equation}
The three cases can be distinguished: $r=1$, $\xi_2=0$ for $|a/c|\ll1$; $r=-1$, $\xi_2=0$ for $|a/c|\gg1$; $r=0$, $\xi_2=\pm1$ for $a/c=\pm1$.

\begin{figure}[tp]
	\centering
	\includegraphics*[width=0.37\linewidth]{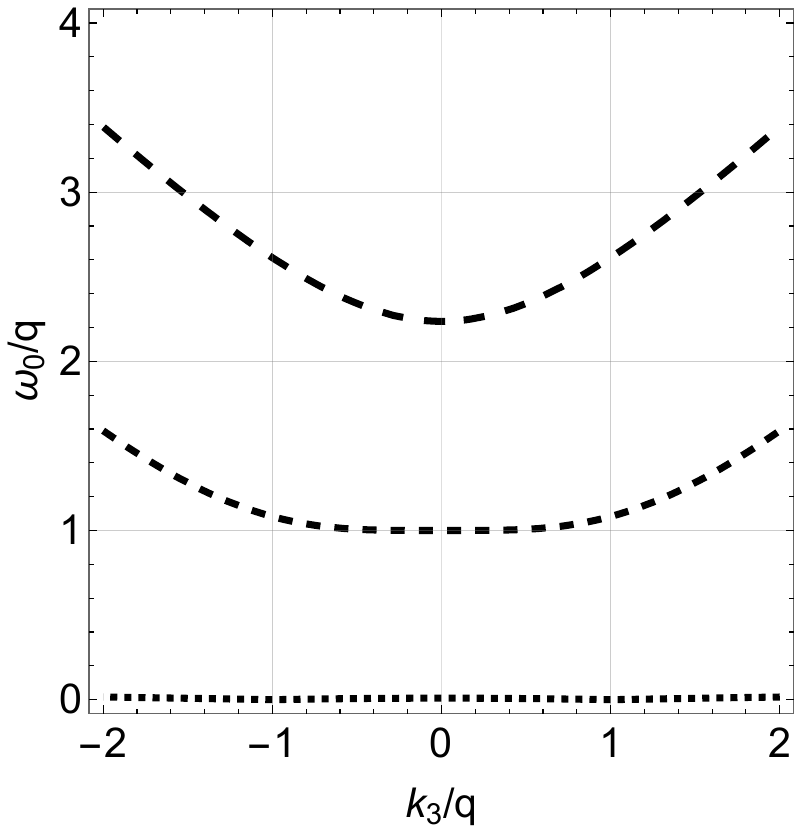}
	\includegraphics*[width=0.6\linewidth]{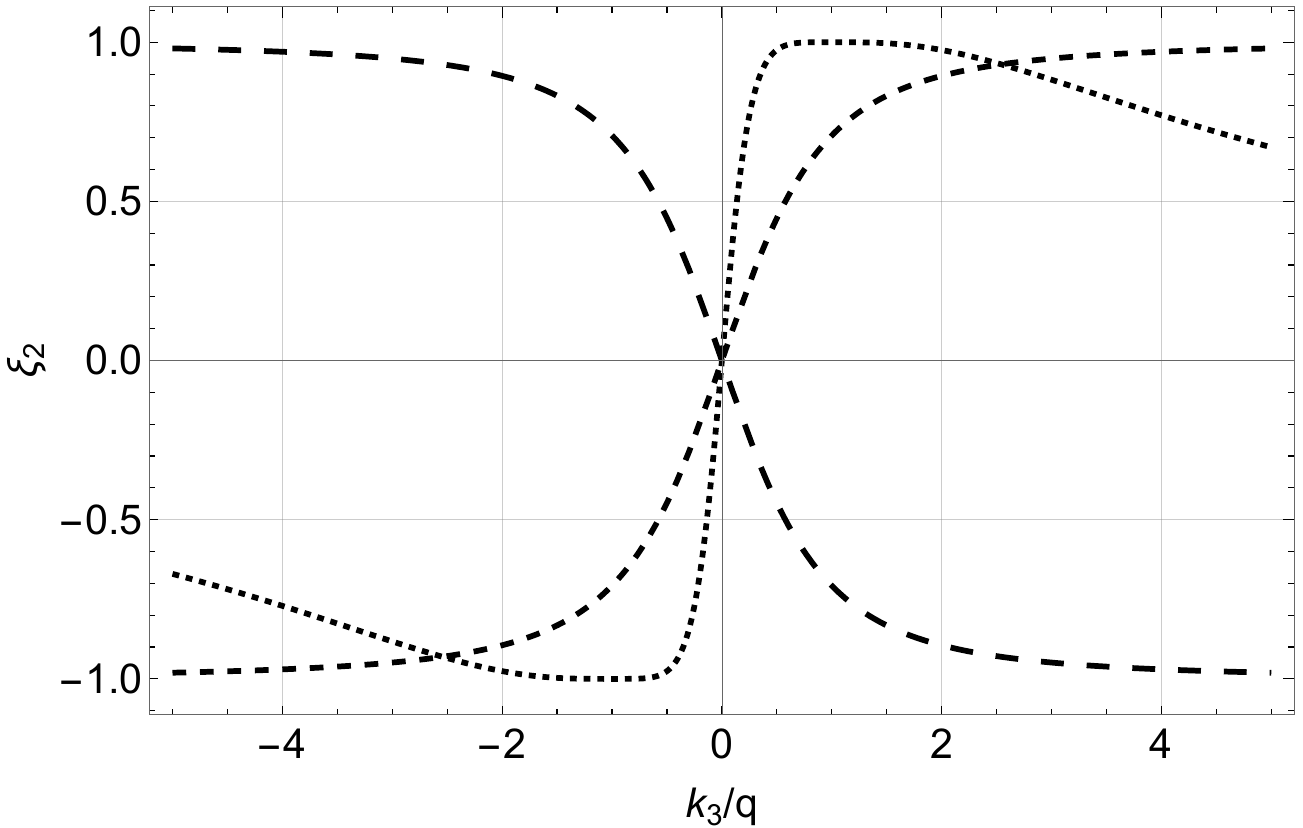}
	\caption{{\footnotesize The dispersion law and the Stokes parameter $\xi_2$ for the different modes of electromagnetic field in the helical wired medium \eqref{diel_permit} in the paraxial limit. The parameters are taken as follows $\pi/2-\alpha=0.01$, $v=1$, $\omega_p=2$, $\e_h=1$, and $q=1$. The second branch of the dispersion law is flat near the point $k_3=0$ for these values of the parameters.}}
	\label{Disp1Flat}
\end{figure}

\begin{figure}[tp]
	\centering
	\includegraphics*[width=0.385\linewidth]{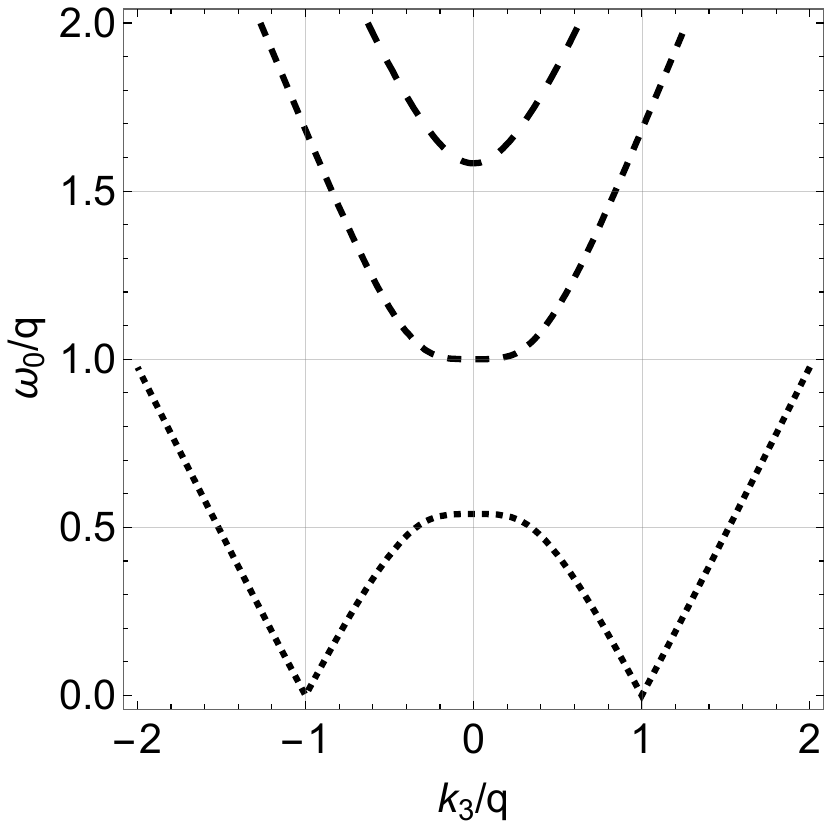}
	\includegraphics*[width=0.6\linewidth]{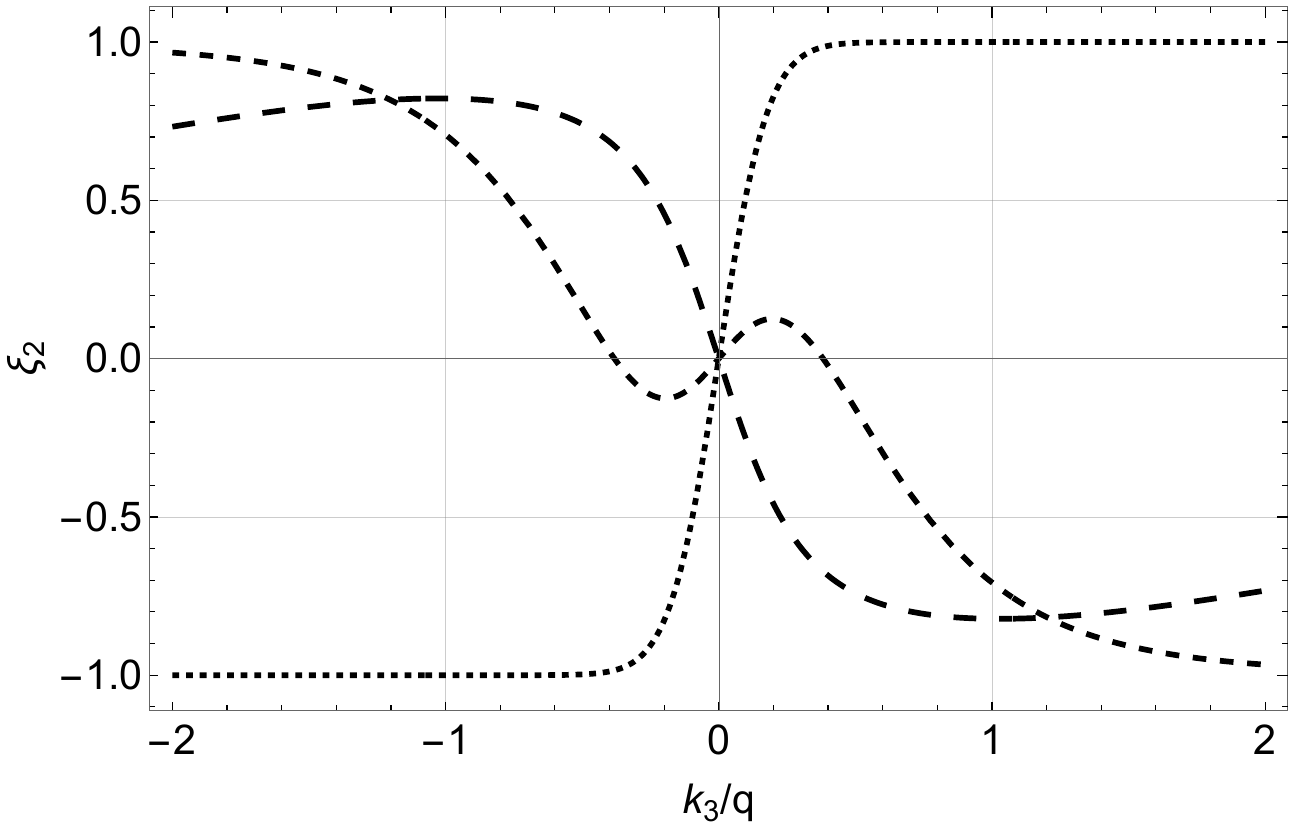}
	\caption{{\footnotesize The dispersion law and the Stokes parameter $\xi_2$ for the different modes of electromagnetic field in the helical wired medium \eqref{diel_permit} in the paraxial limit. The parameters are taken as follows: $\alpha=0.28\pi$, $v=2.65$, $\omega_p=1.34$, $\e_h=1$, and $q=1$. The first and second branches of the dispersion law are flat near the point $k_3=0$ for these values of parameters.}}
	\label{DispBothFlat}
\end{figure}

Then we obtain $a/c\approx\pm1$ for the branches of the dispersion law $\omega^2_0\approx\s^2_\pm$, respectively. As for the electromagnetic field mode with the dispersion law $\omega_0^2\approx\s^2_p$, the Stokes parameters have the form \eqref{xi_plasmon} with
\begin{equation}
    c=k_3^2+q^2-(\omega_p^2+v^2k_3^2)\cos^2\al.
\end{equation}
Whence we have $|a/c|\ll 1$ for $|k_3|\gg |q|$ and $|1-v^2\cos^2\al|k^2_3\gg q^2$. Thus, the modes of electromagnetic field corresponding to the cholesteric branches of the dispersion law, $\sigma_\pm$, possess a circular polarization -- the right-hand polarization for the sign ``$+$'' and the left-hand polarization for the sign ``$-$'' -- exactly as in cholesterics. The electromagnetic field mode corresponding to the polariton branch, $\sigma_p$, possess a linear polarization when $|a/c|\ll 1$. The behavior of the Stokes parameters $\xi_n(k_3)$ is shown in Figs. \ref{DispAsymp}-\ref{DispBothFlat} for the exact branches of the dispersion law $\omega_0(k_3)$.

As in helical media without spatial dispersion \cite{deGennes1993,Yang2006,Belyakov2019,KK2022,Belyakov1982,Lakhtakia1995}, the dispersion law for the model at issue has a chiral forbidden band, i.e., the energy region characterized by a selective reflection of electromagnetic waves with different circular polarizations. Let us estimate the width of this band gap. To this aim, we find the values of $\omega_0(k_3)$ at zero momentum
\begin{equation}
\begin{split}
    \omega_{01} &=\sqrt{\frac{1}{2}\big(q^2+\omega_p^2-\sqrt{q^4+\omega_p^4-2q^2\omega_p^2\cos(2\alpha)}\big)} \leq \min(|q|,\omega_p),\\
    \omega_{02} &=|q|,\\
    \omega_{03} &=\sqrt{\frac{1}{2}\big(q^2+\omega_p^2+\sqrt{q^4+\omega_p^4-2q^2\omega_p^2\cos(2\alpha)}\big)} \geq \max(|q|,\omega_p).
\end{split}
\end{equation}
These energy values obey the inequalities: $\omega_{01}\leqslant \omega_{02} \leqslant \omega_{03}$. We call the branches of the dispersion law with the points $\omega_{01}$, $\omega_{02}$ and $\omega_{03}$ the first, second, and third branches, respectively. The first and second branches can change their shape depending on the value of the plasma frequency $\omega_p$. Namely, the second branch changes its form from $v$-shaped to $w$-shaped provided
\begin{equation}\label{branch_point}
    \omega_0^4(2+v^2\cos^2\al) +\omega_0^2(2q^2(1-v^2\cos^2\al)-\omega_p^2(1+\cos^2\al)) -q^2(2\omega_p^2-v^2q^2)\cos^2\al=0
\end{equation}
at the point $k_3=0$, where $\omega_0=\omega_{02}$ should be substituted. Therefore, the second branch is $v$-shaped for
\begin{equation}
    \omega_p>\omega_{p2}^c=\frac{2|q|}{\sqrt{1+3\cos^2\al}},
\end{equation}
and it is $w$-shaped for $\omega_p<\omega_{p2}^c$. When $\omega_p=\omega_{p2}^c$, the dependence of energy on momentum on the second branch of the dispersion law has the form (see Fig. \ref{Disp1Flat})
\begin{equation}\label{disp_degen}
    \omega_0=\omega_{02}+h k_3^4
\end{equation}
in the vicinity of the point $k_3=0$, where $h$ is some constant. In this case, the density of photon states in the neighborhood of the point $\omega_0=\omega_{02}$ rapidly increases.

\begin{figure}[tp]
\begin{subfigure}{0.28\linewidth}
	\includegraphics*[width=1\linewidth]{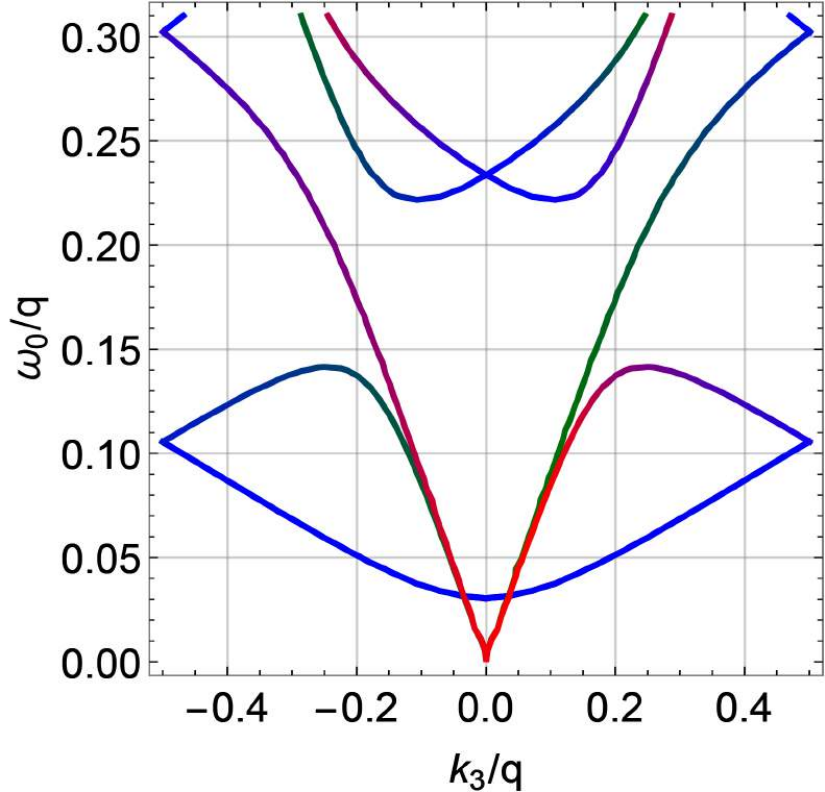}
	\caption{\label{aDisp_exp}}
\end{subfigure}
\begin{subfigure}{0.425\linewidth}
	\includegraphics*[width=1\linewidth]{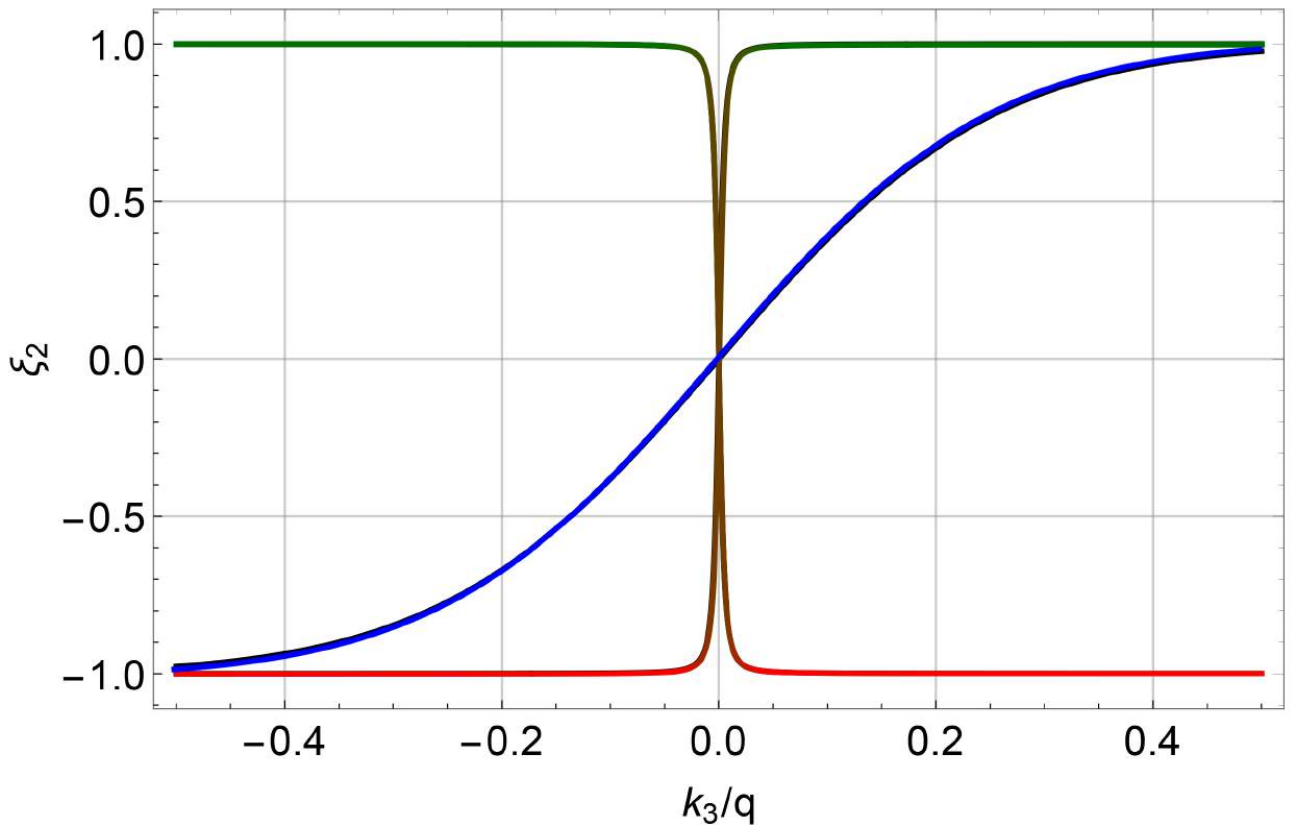}
	\caption{\label{bXi2_exp}}
\end{subfigure}
\begin{subfigure}{0.285\linewidth}
	\includegraphics*[width=1\linewidth]{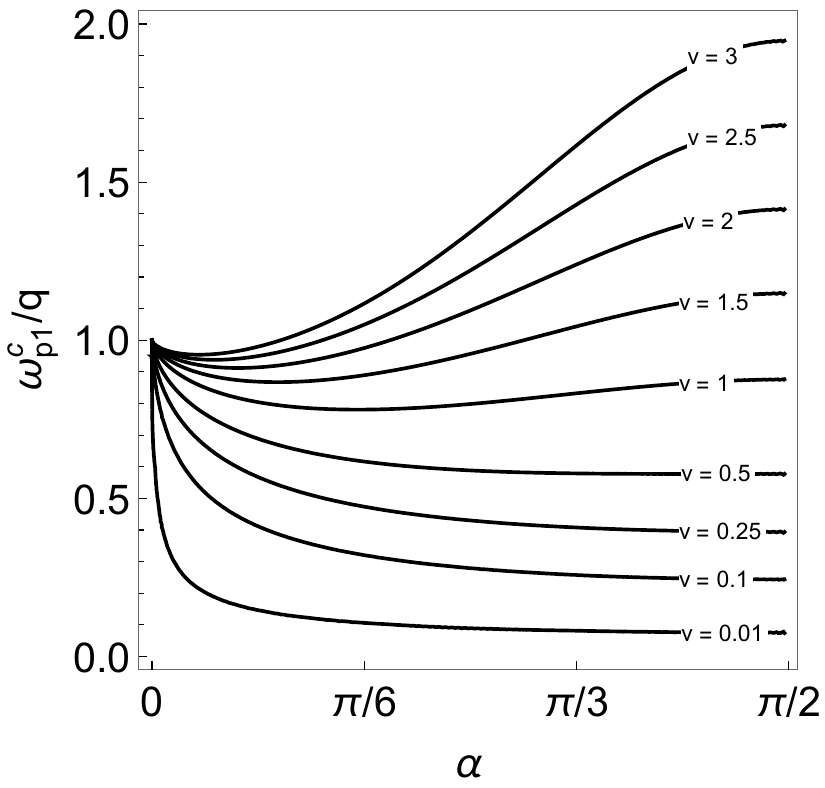}
	\caption{\label{cOmegaP1C}}
\end{subfigure}
\caption{{\footnotesize The plots $(a)$ and $(b)$: The dispersion law reduced to the first Brillouin zone and the Stokes parameter $\xi_2$ for the different modes of electromagnetic field in the helical wired medium \eqref{diel_permit} in the paraxial limit. The parameters are taken as follows: $\alpha=\arctan(2\pi\times 3.3/4.4)$, $v=1$, $\omega_p=0.15$, $\e_h=1$, and $q=1$. The colors of the curves are chosen in agreement with the rule described in Fig. \ref{DispAsymp}. The plot $(c)$: The dependence $\omega^c_{1p}(\alpha,v)$.}}
\label{OmegaP1C}
\end{figure}

The first branch can also change its shape: this branch of the dispersion law acquires an additional dip in the vicinity of the point $k_3=0$ for sufficiently small $\omega_p$ (see Figs. \ref{DispAsymp}, \ref{DispBothFlat}). The critical value $\omega_{p1}^c$ where this transition occurs is determined by equation \eqref{branch_point} with $\omega_0=\omega_{01}$. The resulting equation on $\omega^c_{p1}$ is exactly solvable but this solution is rather huge. The dependence of $\omega_{p1}^c$ on $\al$ and $v$ is shown in Fig. \ref{OmegaP1C}. For $\omega_p=\omega^c_{p1}$, the dependence of energy on momentum for the first branch of the dispersion law has the form \eqref{disp_degen} near the point $k_3=0$, where one must substitute $\omega_{02}\rightarrow\omega_{01}$ and take another constant $h$. In this case, the density of photon states rapidly increases in the vicinity of the point $\omega_0=\omega_{01}$. It is possible to choose the parameters $\al$ and $v$ so that both the first and second branches of the dispersion law are flat near the point $k_3=0$. Then
\begin{equation}
    v^2=\frac{16}{\sin\al\cos^2\al}\frac{(1+\cos^2\al)\sqrt{25+39\cos^2\al}-(5+9\cos^2\al)\sin\al}{(\sqrt{25+39\cos^2\al}-3\sin\al)^2}.
\end{equation}
The dispersion law in this case is shown in Fig. \ref{DispBothFlat}.

\begin{figure}[tp]
	\centering
	\includegraphics*[width=0.97\linewidth]{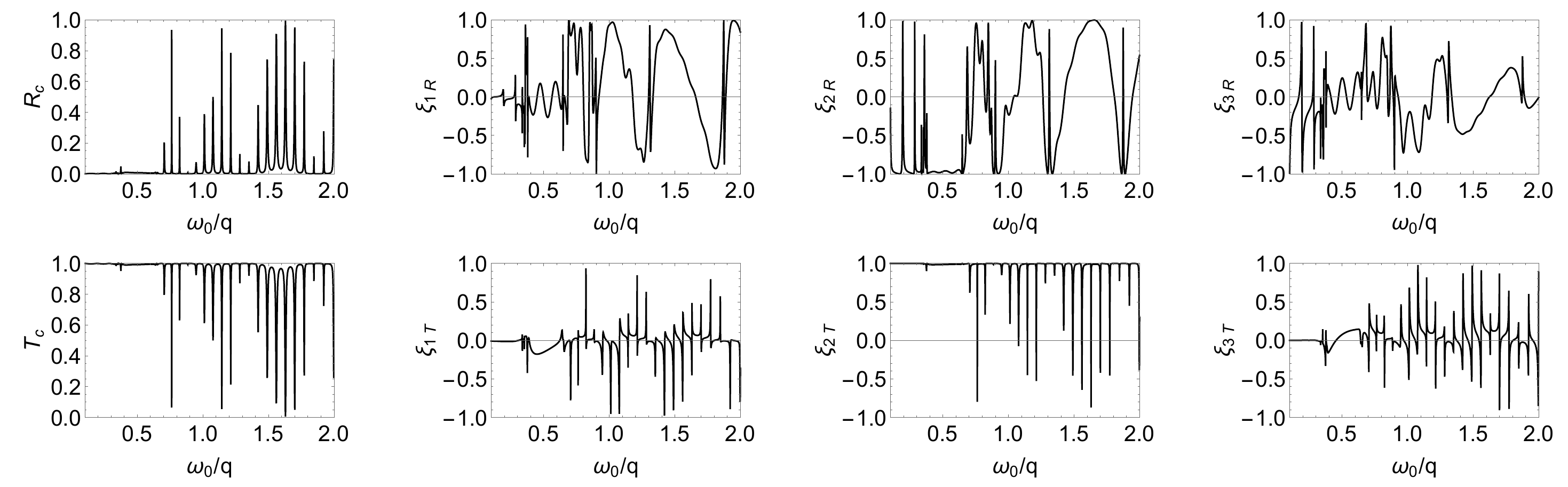}
	\includegraphics*[width=0.97\linewidth]{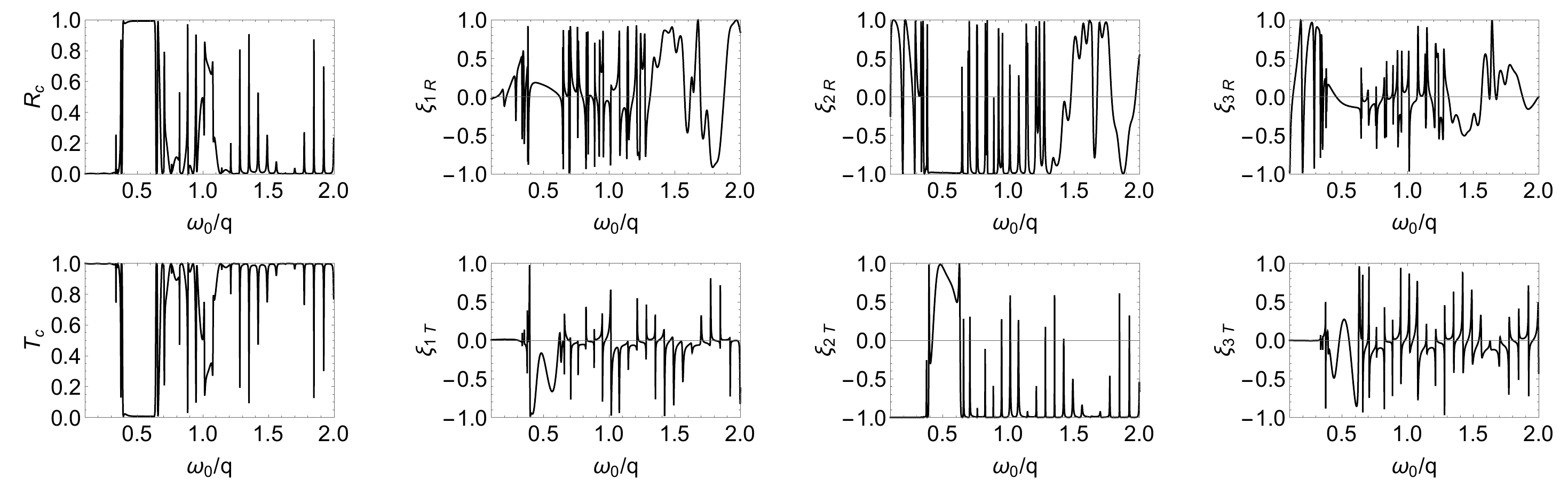}
	\caption{{\footnotesize The plots of the reflection coefficients $R_c$, the transmission coefficients $T_c$, and the Stokes parameters of the reflected and transmitted waves. The incident plane wave has either the right-hand polarization (the first and second lines) or the left-hand one (the third and fourth lines). The parameters are taken as follows: $\alpha=\pi/4$, $v=1$, $\omega_p=0.5$, $\e_0=\e_h=1$, the slab width $L=10\pi$, and $q=1$. The dispersion law of the electromagnetic field modes is shown in Fig. \ref{DispAsymp}.}}
	\label{Scattering1}
\end{figure}

\begin{figure}[tp]
	\centering
	\includegraphics*[width=0.97\linewidth]{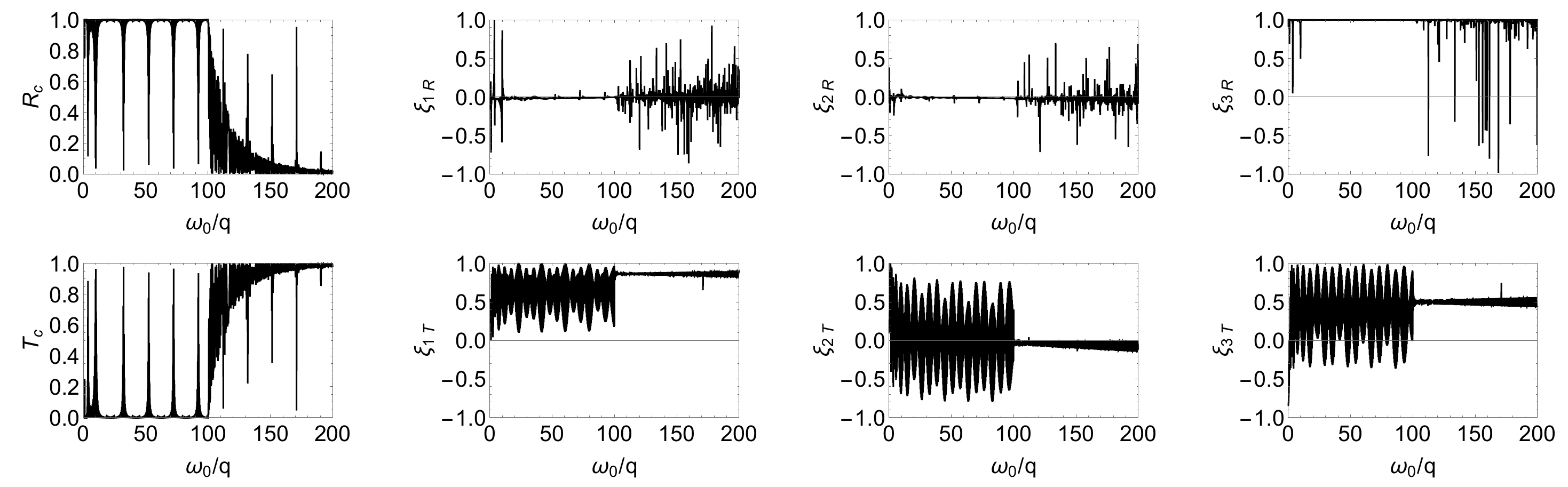}
	\includegraphics*[width=0.97\linewidth]{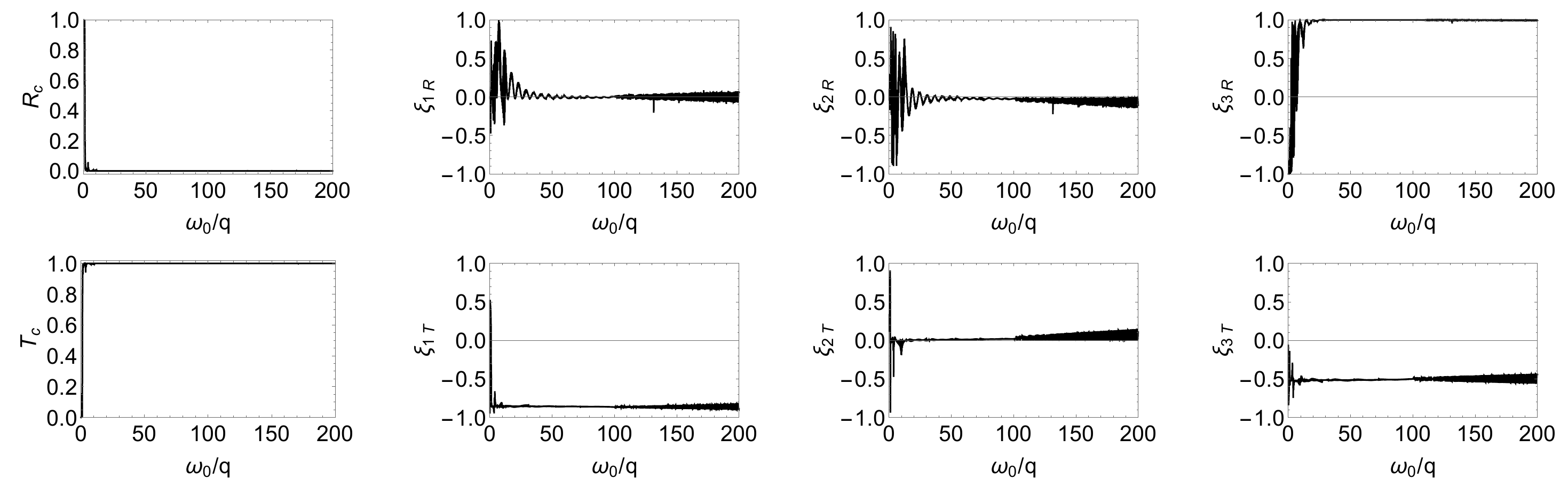}
	\caption{{\footnotesize The plots of the reflection coefficients $R_c$, the transmission coefficients $T_c$, and the Stokes parameters of the reflected and transmitted waves. The incident plane wave has the linear polarization with either the Stokes parameter $\xi_3=1$ (the first and second lines) or the Stokes parameter $\xi_3=-1$ (the third and fourth lines). The parameters are taken as follows: $\pi/2-\alpha=0.01$, $v=1$, $\omega_p=100$, $\e_0=\e_h=1$, the slab width $L=61\pi/6$, and $q=1$. The dispersion law of the electromagnetic field modes is shown in Fig. \ref{DispWKB}.}}
	\label{Scattering2}
\end{figure}

\begin{figure}[tp]
	\centering
	\includegraphics*[width=0.97\linewidth]{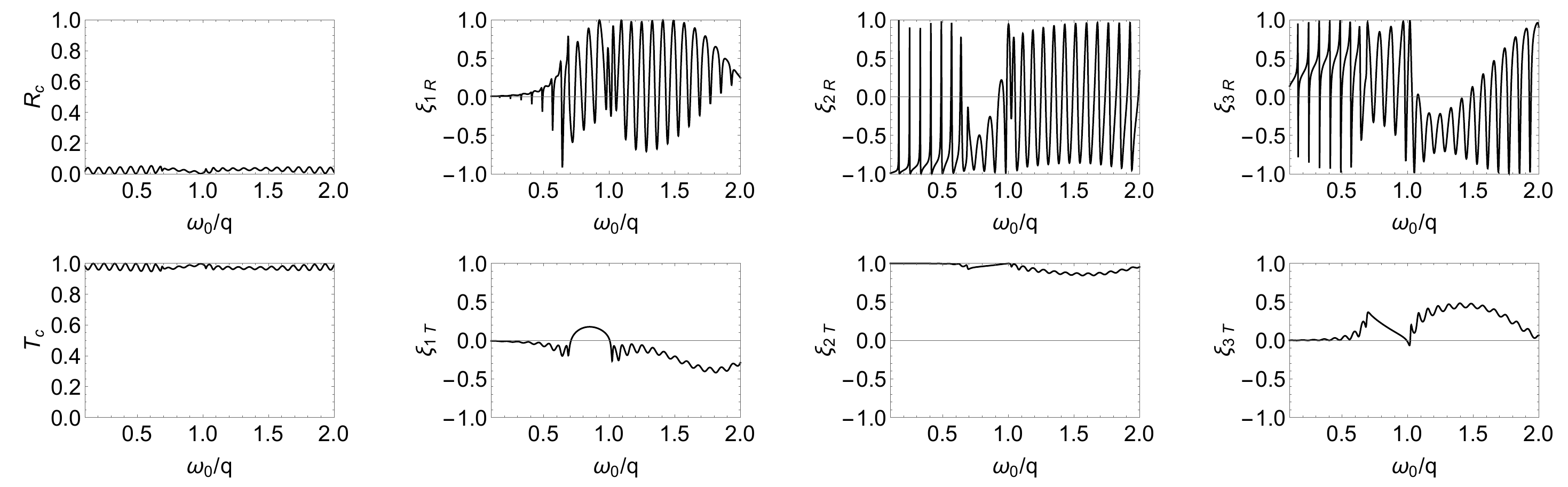}
	\includegraphics*[width=0.97\linewidth]{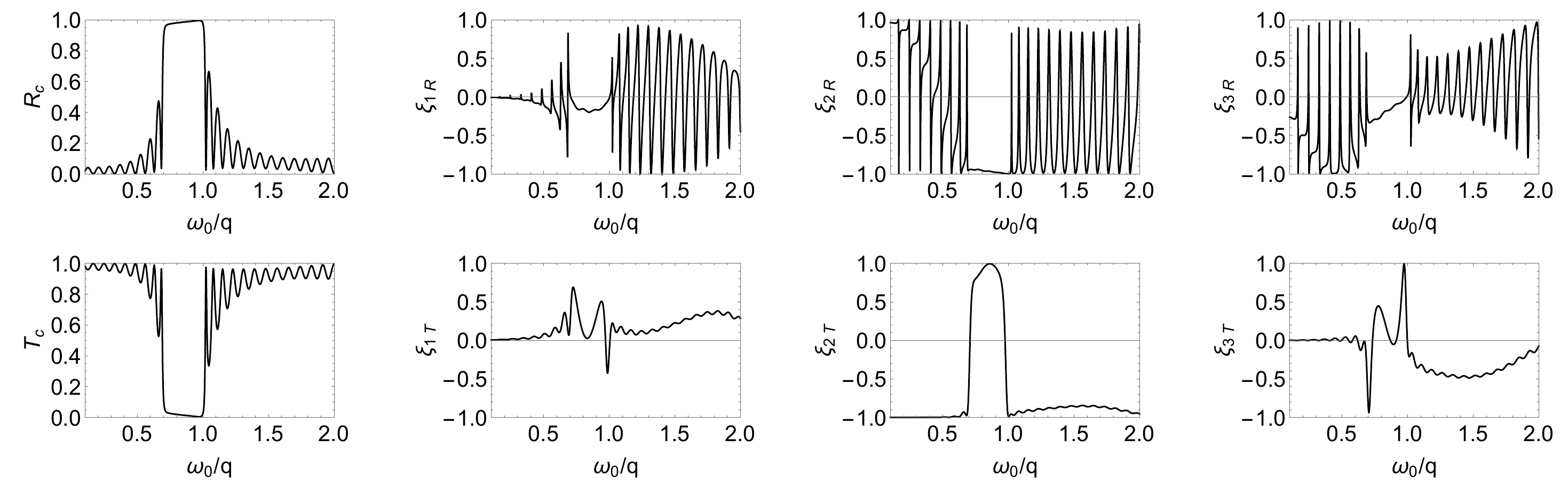}
	\caption{{\footnotesize The same as in Fig. \ref{Scattering1} but the parameters are taken as follows: $\alpha=\pi/4$, $v=1$, $\omega_p=15$. The dispersion law of electromagnetic field modes is shown in Fig. \ref{DispColored}.}}
	\label{Scattering2-5}
\end{figure}

In the case
\begin{equation}\label{simple_gap}
    \omega_p\geqslant\omega_{p1}^c,\qquad\omega_p\geqslant \omega_{p2}^c,
\end{equation}
the energy range $\omega_0 \in (\omega_{01},\omega_{02})$ corresponds to the chiral forbidden band. Indeed, in this energy region, the energies close to the asymptotic branch $\omega_0 \approx \sigma_+$ are realized for the electromagnetic waves propagating in the positive direction, i.e., with positive group velocity (see Figs. \ref{DispAsymp}, \ref{DispColored}). These waves possess a right-hand circular polarization. Similarly, only the left-hand circularly polarized electromagnetic waves with energies $\omega_0 \approx \sigma_-$ propagate in the opposite direction, i.e., with negative group velocity. Thus, for the energies belonging to the chiral forbidden band, only the electromagnetic waves whose chirality coincides with chirality of the helical wired medium pass through it. In other words, the chirality of transmitted electromagnetic waves coincides with the sign of $q$. When the sign of $q$ is flipped, the cholesteric asymptotics interchange, $\sigma_{\pm}\rightarrow \sigma_{\mp}$, and so does the helicity of the electromagnetic waves transmitted by this medium in the chiral forbidden band. The plots of the reflection and transmission coefficients and the Stokes parameters for the transmitted and reflected waves are presented in Figs. \ref{Scattering1}-\ref{Scattering3}. It is assumed that the electromagnetic wave falls normally on the plane-parallel slab of thickness $L$ from the region $z<0$. The plate is made of the helical wired medium with the helical axis $z$ and is placed perpendicular to the $z$ axis. The plots of the transmission and reflection coefficients clearly show the presence of the chiral forbidden band for electromagnetic waves in a helical wired medium. If the inequalities \eqref{simple_gap} are violated as, for example, in Fig. \ref{DispAsymp}, then the expressions for $\omega_{01}$ and $\omega_{02}$ can be used to estimate the positions of the forbidden band edges. In the case $\omega_p\ll|q|$, the two chiral forbidden bands appear: for $\omega_0\approx\omega_p$ and for $\omega_0\in(|q|,\sqrt{q^2+\omega^2_p\sin^2\al})$ (see Figs. \ref{DispAsymp}, \ref{Scattering1}).

In order to verify the agreement of the constructed effective model with the numerical simulations and the results of other papers, we present in Fig. \ref{OmegaP1C} the plots of the dispersion law reduced to the first Brillouin zone and the Stokes parameter $\xi_2$ for the helix angle $\al$ taken from the works \cite{Wu2010,LiWongChan2015}. We see good agreement of the dispersion law with the results obtained by the use of the semianalytical model and by the numerical simulations (compare Fig. \ref{OmegaP1C} with Fig. 2.b in \cite{Wu2010} and Figs. 5.2.b, 5.3 in \cite{LiWongChan2015}).

To characterize the energy flux density carried by the electromagnetic field in the helical wired medium, we find the direction of the Poynting vector, $\bs\Pi$, for the above solutions of the Maxwell equations. The Poynting vector must, of course, be calculated with the aid of the real field strengths of the corresponding modes
\begin{equation}\label{Point_defn}
    \mathbf{E}=-k_0\im \mathbf{A},\qquad \mathbf{H}=\re\rot\mathbf{A},\qquad \bs\Pi=[\mathbf{E},\mathbf{H}].
\end{equation}
Notice that the plasmon field $\Psi$ also carries the energy and momentum along the conducting wires. We will be interested only in the part of the energy-momentum that is carried by the electromagnetic field. Then the components of the Poynting vector can be written as
\begin{equation}\label{Poynting}
\begin{split}
    \Pi_+&=\frac{k_0 \omega_p^2 e^{iqz}\sin(k_3 z)\sin(2\alpha) }{4(\omega_0^2-\sigma_+^2)(\omega_0^2-\sigma_-^2)}\big[iq(\omega_0^2+k_3^2-q^2)\cos(k_3z)-k_3(\omega_0^2-k_3^2+q^2)\sin (k_3z) \big],\\
    \Pi_3 &= \frac{k_3 k_0 \omega_0^2 \omega_p^2 \sin^2\alpha}{4(\omega_0^2-\sigma_+^2)(\omega_0^2-\sigma_-^2)} \Big[ \frac{(\omega_0^2-k_3^2+q^2)^2+4q^2(k_3^2-q^2)}{(\omega_0^2-\sigma_+^2)(\omega_0^2-\sigma_-^2)} - \cos(2k_3 z)\Big],
\end{split}
\end{equation}
where, we recall, $\Pi_+=\Pi_1+i\Pi_2$. To shorten the notation, we consider the Poynting vector at $t=0$ hereinafter in this section. To restore the expression for the Poynting vector at an arbitrary instant of time $t$, one just has to replace $k_3z\rightarrow k_3z-k_0t$.

To describe the integral curves of the Poynting vector, we introduce the Frenet frame $\{\bs\tau,\mathbf{n},\mathbf{b}\}$, where $\bs\tau$ is the tangent vector to the conducting wires forming the helices, $\mathbf{n}$ is the normal vector to these helices, and $\mathbf{b}$ is the binormal vector. The normal and binormal vectors are written as
\begin{equation}
    \mathbf{n}=\frac{\partial_z\bs\tau}{q\sin\alpha}=[\spe_3,\mathbf{d}],\qquad\mathbf{b}=[\bs\tau,\mathbf{n}]=\sin\al\spe_3-\cos\al\mathbf{d}.
\end{equation}
The helically symmetric vectors $\{\bs\tau,\mathbf{n},\mathbf{b}\}$ constitute a right-handed orthonormal triple. The components of the Pointing vector in the Frenet frame can be cast into the form
\begin{equation}\label{Sprojection}
\begin{split}
	(\bs\Pi\bs\tau)=& \frac{k_0k_3 \omega_p^2\sin^2\alpha\cos\alpha }{4(\omega_0^2-\sigma_+^2)(\omega_0^2-\sigma_-^2)}
    \Big[\frac{4q^2\omega_0^4 +(k_3^2-q^2)(\omega_0^2-k_3^2+q^2)^2}{(\omega_0^2-\sigma_+^2)(\omega_0^2-\sigma_-^2)}-(k_3^2-q^2)\cos(2k_3z) \Big],\\
	(\bs\Pi\mathbf{n})=& \frac{k_0 \omega_p^2q\sin(2\al)
    (\omega_0^2+k_3^2-q^2)}{8(\omega_0^2-\sigma_+^2)(\omega_0^2-\sigma_-^2)}\sin(2k_3 z),\\
	(\bs\Pi\mathbf{b})=&
    \frac{k_0 k_3 \omega_p^2\sin\alpha}{4(\omega_0^2-\sigma_+^2)(\omega_0^2-\sigma_-^2)} \Big[\omega_0^2-k_3^2+q^2+\frac{4q^2\omega_0^4+(k_3^2-q^2)(\omega_0^2-k_3^2+q^2)^2}{(\omega_0^2-\sigma_+^2)(\omega_0^2-\sigma_-^2)} \sin^2\al-\\
    &-(\omega_0^2 -(k_3^2-q^2)\cos^2\alpha)\cos(2k_3z)\Big].
\end{split}
\end{equation}
In increasing $z$, the components of the Poynting vector in the Frenet basis oscillate with frequency $2k_3$ near their mean values. The mean values of the components of $\bs\Pi$ in the Frenet basis are obtained by substituting $\cos(2k_3z)\rightarrow0$, $\sin(2k_3z)\rightarrow0$ in expressions \eqref{Sprojection}. The Poynting vector averaged in this way is given by
\begin{equation}\label{Point_ave}
    \lan\bs\Pi\ran=\lan(\bs\Pi\bs\tau)\ran\bs\tau+\lan(\bs\Pi\mathbf{b})\ran\mathbf{b},
\end{equation}
where the averaged projections of the vector $\bs\Pi$ do not depend on $z$. It is clear that expression \eqref{Point_ave} also describes the time-averaged Poynting vector. In increasing $z$, the Frenet frame rotates with frequency $q$. Therefore, the evolution of the Poynting vector with increasing $z$ looks as beats (see Fig. \eqref{PoyntingCurves}). It should be noted, however, that in the energy region
\begin{equation}\label{energy_range}
    \omega_0\in\big(\sqrt{|k_3^2-q^2|},\sqrt{k_3^2+q^2}\big)
\end{equation}
there exist such values of $z$ that $\Pi_3$ vanishes. As a result, in this energy range the instantaneous integral curves of the Poynting vector consist of two lines lying in the planes $z=z_1$ and $z=z_2$, where $z_{1,2}$ are the two neighboring zeros of $\Pi_3(z)$. These two lines are smoothly connected by an arc of the length of order $1/k_3$ (see Fig. \ref{PoyntingCurves}).

In the vicinity of the cholesteric asymptotics, $\omega_0=\s_\pm$, the Poynting vector is proportional to
\begin{equation}
\begin{split}
    \bs\Pi \sim&\, \spe_3 -\Big\{\frac{\cos\al}{\sin^2\al}\big[2q \mp(k_3\pm q)\sin^2\al\cos(2k_3z)\big]\bs\tau\\
    &\pm 2q\cot\al\sin(2k_3z)\mathbf{n} + \big[2q-(k_3 \pm q)\sin^2\al\big]\frac{\cos(2k_3z)}{\sin\al}\mathbf{b} \Big\}\frac{\omega_0-\s_\pm}{q\s_\pm}+\cdots.
\end{split}
\end{equation}
The expression for $\bs\Pi$ in the vicinity of the polariton asymptotics, $\omega_0=\s_p$, is rather cumbersome and we present here only the leading contribution assuming that $|k_3|\gg|q|$. Then the Poynting vector is proportional to
\begin{equation}
    \bs\Pi\sim k_3^2\tan\al\bs\tau +(\omega_p^2-(1-v^2)k_3^2)\mathbf{b}.
\end{equation}
In particular, when $v=1$ and $k_3^2\tan\al\gg\omega^2_p$, we derive that
\begin{equation}
    \bs\Pi\sim \bs\tau,
\end{equation}
on this branch of the dispersion law. Thus, we see that the electromagnetic field modes close to the cholesteric asymptotics propagate along the $z$ axis, while the polariton mode, $\omega_0\approx\s_p$, propagates along the conducting wires at large momenta.

Let us also consider the behavior of the Poynting vector in the shortwave regime \eqref{wkb_regime}. In this case,
\begin{equation}\label{S_wkb_parax}
    \frac{\Pi_+}{\Pi_3}\approx e^{iqz}\cot\al\frac{k_3^2-\omega_0^2}{\omega_0^2}.
\end{equation}
Therefore, we have
\begin{equation}\label{Point_wkbpar1}
    \bs\Pi\sim \spe_3,
\end{equation}
for the branch of the dispersion law $\omega_0^2\approx k_3^2$. In order to find the direction of propagation of the electromagnetic wave energy density for the rest two modes, we have to take the expressions on the first line of \eqref{wkb_disp} and substitute them into \eqref{S_wkb_parax}. In particular, for $v=1$ we find that
\begin{equation}\label{Point_wkbpar23}
    \bs\Pi\sim\bs\tau,\qquad\bs\Pi\sim\omega_0^2\sin\al\spe_3-\omega_p^2\cos\al\mathbf{d},
\end{equation}
for the branches of the dispersion law $\omega_0^2=k_3^2\cos^2\al$ and $\omega_0^2=\omega_p^2+k_3^2$, respectively.

\subsection{Particular cases}

Consider the particular cases of the dispersion law \eqref{ParaxDisp} for certain values of $\al$. The asymptotics  \eqref{ParaxDispAsym} become exact solutions of the dispersion law \eqref{ParaxDisp} in the case $\alpha=0$. This limit describes degeneracy of helices into straight lines directed along the $z$ axis. For $\al=0$, the matrix \eqref{ParaxFourierCoef} is diagonalized and the solution \eqref{ParaxSol} is invalid. Such wired media were profoundly studied in the literature \cite{Belov2003,Maslovski2009,Tyukhtin2014,Yakovlev2020,Silverinha2006}. In the present paper, we investigate the case of small but nonzero $\alpha$. For small $\al$, the branches of the dispersion law and the polarization properties of the corresponding modes are described with good accuracy by the asymptotics \eqref{ParaxDispAsym}, \eqref{xi_plasmon}. The Poynting vector for modes with $\omega_0\approx\s_\pm$ is proportional to
\begin{equation}
    \bs\Pi\sim\spe_3+\frac{\al\omega_p^2}{2}\frac{\mathbf{n}+\mathbf{d}(1-\cos(2k_3z))}{\omega_p^2+v^2k_3^2-(k_3\mp q)^2}.
\end{equation}
As far as the polariton mode of electromagnetic field, $\omega_0\approx\s_p$, is concerned, the Poynting vector becomes
\begin{equation}
    \bs\Pi\sim-k_3\big(\omega_p^2-(1-v^2)k_3^2+q^2\big)\big(1-\cos(2k_3z)\big)\mathbf{d} +q\big(\omega_p^2+(1+v^2)k_3^2-q^2\big) \sin(2k_3z)\mathbf{n},
\end{equation}
in leading order. We see that in this case the electromagnetic wave modes corresponding to the cholesteric branches of the dispersion law propagate approximately along the $z$ axis, whereas the polariton mode propagates in the plane perpendicular to this axis.

\begin{figure}[tp]
	\centering
	\includegraphics*[width=0.97\linewidth]{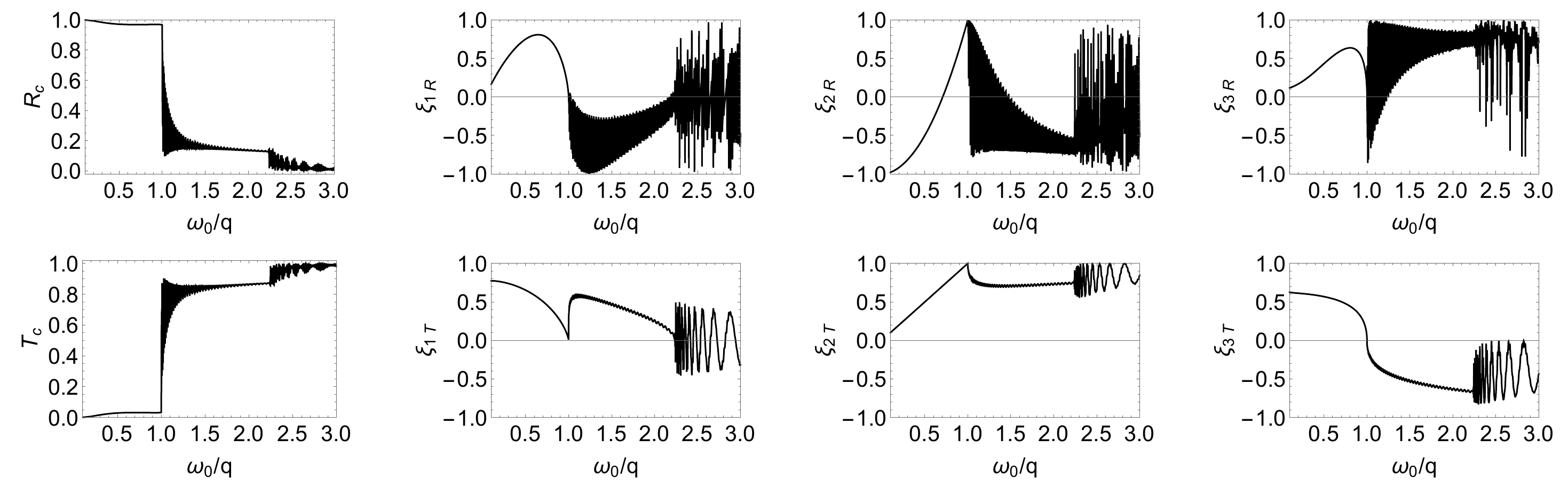}
	\includegraphics*[width=0.97\linewidth]{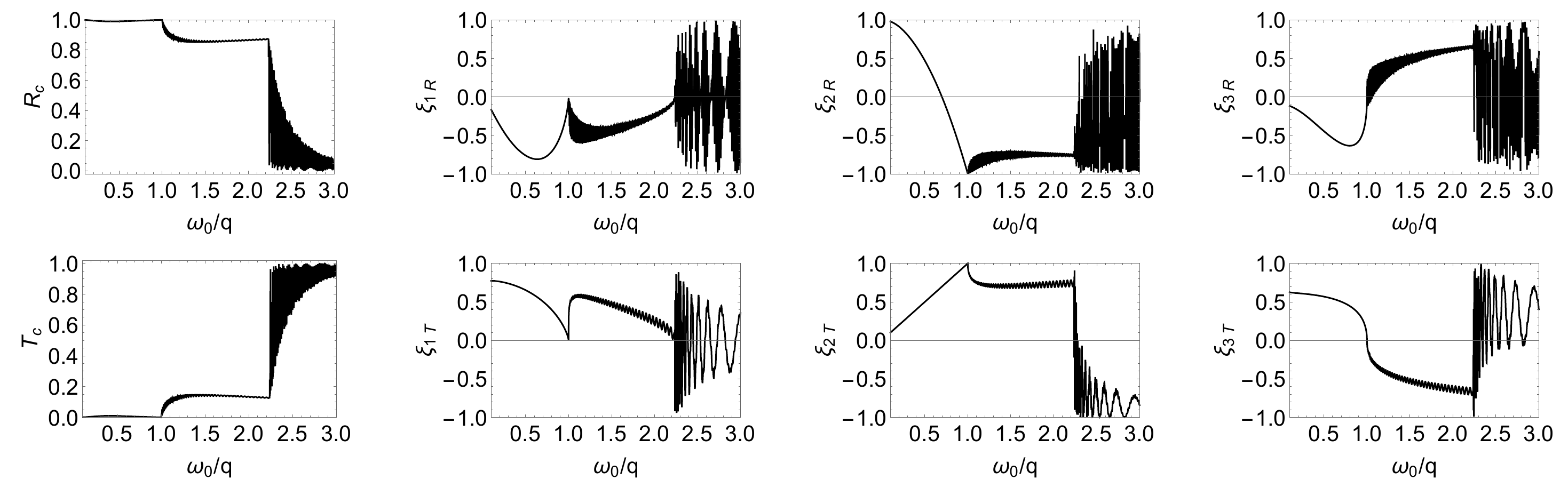}
	\caption{{\footnotesize The same as in Fig. \ref{Scattering1} but the parameters are taken as follows $\pi/2-\alpha=0.01$, $v=1$, $\omega_p=2$, and $L=40\pi$. The dispersion law of the electromagnetic field modes is shown in Fig. \ref{Disp1Flat}. There is the total forbidden band at $\omega_0<1$ for these values of parameters.}}
	\label{Scattering2-6}
\end{figure}

Another important case corresponds to the limit $\alpha=\pi/2$. The physical realization of this metamaterial is arranged similarly to cholesterics: the metamaterial consists of thin layers, every such a layer comprises of parallel conducting wires directed along the director $\mathbf{d}(z)$, and all the layers are orthogonal to the vector $\mathbf{e}_3$. As one moves from layer to layer, the director $\mathbf{d}(z)$ rotates according to formula \eqref{xi_tau}. In the leading order in $(\al-\pi/2)$, the dispersion law turns into
\begin{equation}\label{CholLikeDisp}
    \omega_0^2(k_3)=q^2+k_3^2 + \frac{\omega_p^2}{2} \pm \frac{1}{2}\sqrt{\omega_p^4+16q^2 k_3^2},\qquad \omega_0^2(k_3)=(\al-\pi/2)^2\frac{(\omega_p^2+v^2k_3^2) (k_3^2-q^2)^2}{\omega_p^2(k_3^2+q^2)+(k_3^2-q^2)^2}.
\end{equation}
One of the three branches of the dispersion law is degenerate and tends to zero. In the regime \eqref{estim_asympt}, the other two branches are well approximated by the asymptotics $\sigma_{\pm}$ and have the respective polarization properties. These branches of the dispersion law resemble the dispersion law of electromagnetic waves in cholesteric liquid crystals \cite{deGennes1993,Yang2006,Belyakov2019,KK2022,Belyakov1982,Lakhtakia1995}. In this case, the vector $\bs\tau$ coincides with the vector $\mathbf{d}$ corresponding to the cholesteric director. The polarization properties of these modes agree with the polarization properties of the electromagnetic field modes in cholesterics \cite{deGennes1993,Yang2006,Belyakov2019,KK2022,Belyakov1982,Lakhtakia1995}. There is, however, an important qualitative difference between helical wired media with $\al\approx\pi/2$ and cholesterics. If we discard the degenerate branch, the admissible photon energies are bounded from below by a positive constant. Indeed, it follows from \eqref{CholLikeDisp} that for the non-generated branches of the dispersion law
\begin{equation}\label{omega_0m}
	\omega_0^2(k_3)\geq\omega_{0m}^2=
\begin{cases}
    \frac{\omega_p^2}{2}\big(1-\frac{\omega_p^2}{8q^2}\big) ,& \text{при} \quad \omega_p<2|q|;\\
    q^2, & \text{при} \quad \omega_p\geqslant 2|q|.
\end{cases}
\end{equation}
If $\omega_p=2|q|$, then the lower nondegenerate branch of the dispersion law becomes flat near the point $k_3=0$, i.e., it has the form \eqref{disp_degen} with $\omega_{02}=|q|$ (see Fig. \ref{Disp1Flat}). As long as the contribution of the degenerate branch to the scattering of electromagnetic waves by such a medium is suppressed, the forbidden band for the electromagnetic waves of any polarization is realized for the energies $\omega_0\in(0,\omega_{0m})$ (see Fig. \ref{Scattering2-6}). In the region of applicability of the shortwave approximation, $|k_3||\sim\omega_p\gg|q|$, the plots of the dispersion law and the Stokes parameter $\xi_2$ are shown in Fig. \ref{DispWKB}, and the respective scattering data are presented in Fig. \ref{Scattering2}. The photon energy domain $\omega_0\in(0,\omega_{0m})=(0,|q|)$ is poorly visible in Fig. \ref{Scattering2} due to the chosen scale. In this energy range, the helical wired medium reflects the electromagnetic waves of any polarization.

In leading order in $(\al-\pi/2)$, the Poynting vector is proportional to
\begin{equation}
    \bs\Pi\sim\spe_3,
\end{equation}
for the electromagnetic field modes corresponding to nondegenerate branches of the dispersion law. In other words, these modes propagate along the $z$ axis. As for the branch of the dispersion law tending to zero as $\al\rightarrow\pi/2$, the Pointing vector is proportional to
\begin{equation}
    \bs\Pi\sim k_3(1-\cos(2k_3z))\mathbf{d} +q\sin(2k_3z)\mathbf{n},
\end{equation}
i.e., this mode of electromagnetic field propagates in the plane perpendicular to the $z$ axis. The averaged Poynting vector for this mode is directed along the conducting wires because $\mathbf{d}\approx\bs\tau$.

\section{Shortwave approximation}\label{Shrt_Wave_Appr}

Let us find a solution to the system of equations \eqref{MaxwellEq2} in the shortwave approximation. We assume that $\omega_0$, $k_\perp$, $\bar{k}_3$, and $\omega_p$ are large parameters and apply the matrix shortwave approximation procedure \cite{BabBuld,Maslov,BagTrif,RMKD18,BKKL21}. For a self-consistent application of the this procedure, we can introduce a formal parameter $\la$ as
\begin{equation}
    \omega_0\rightarrow\la\omega_0,\qquad k_\perp\rightarrow\la k_\perp,\qquad \bar{k}_3\rightarrow\la\bar{k}_3,\qquad\omega_p\rightarrow\la\omega_p.
\end{equation}
Then we expand all the quantities in inverse powers of this parameter and put $\la=1$ in the final answer. Henceforth, we will not write explicitly the parameter $\la$.

According to the procedure developed in \cite{BabBuld,Maslov,BagTrif,RMKD18,BKKL21}, we seek for a solution to equation \eqref{MaxwellEq2} in the form
\begin{equation}\label{sh_w_expn}
    W=[L_0(z)+L_{-1}(z)+L_{-2}(z)+\cdots] e^{iS(z)},
\end{equation}
where the index at $L_k(z)$ denotes the order in powers of $\la$. It is also assumed that the function $S(z)$ and its derivatives are of first order in $\la$. Notice that the index at $M_k$ in equation \eqref{MaxwellEq2} coincides with the order of this matrix in $\la$. Substituting the expansion \eqref{sh_w_expn} into \eqref{MaxwellEq2} and collecting the terms of the same order in $\la$, we obtain
\begin{equation}\label{first_two_orders}
\begin{split}
    (k_3^2M_0 +k_3M_1 +M_2) L_0&=0,\\
    (k_3^2M_0 +k_3M_1 +M_2) L_{-1}&=i\big[\frac12(M_1L_0)' +\frac12 M_1L_0' +2k_3 M_0L_0' +k_3'M_0L_0 +k_3 M_0'L_0 \big],
\end{split}
\end{equation}
in the first two orders, where $k_3:=S'(z)$. Let $L_0(z)$ be the unique solution of the first equation up to multiplication by some function. Then multiplying the second equation from the left by $L^\dag(z)$, we come to
\begin{equation}
    \frac12 L_0^\dag(M_1L_0)' +\frac12 L_0^\dag M_1L_0' +2k_3 L_0^\dag M_0L_0' +k_3'L_0^\dag M_0L_0 +k_3 L_0^\dag M_0'L_0=0.
\end{equation}
Separating the real and imaginary parts in this equation, we have
\begin{equation}\label{consist_eq}
\begin{split}
    (k_3 L^\dag_0M_0L_0 +\frac12 L_0^\dag M_1L_0)'&=0,\\
    2k_3(L^\dag_0M_0L'_0 -L'^\dag_0M_0L_0) +L_0^\dag M_1L'_0-L'^\dag_0 M_1L_0&=0,
\end{split}
\end{equation}
where $k_3$ is assumed to be real. The fulfillment of these equations is a necessary and sufficient solvability condition for the second equation in \eqref{first_two_orders}.

\begin{figure}[tp]
	\centering
	\includegraphics*[width=0.97\linewidth]{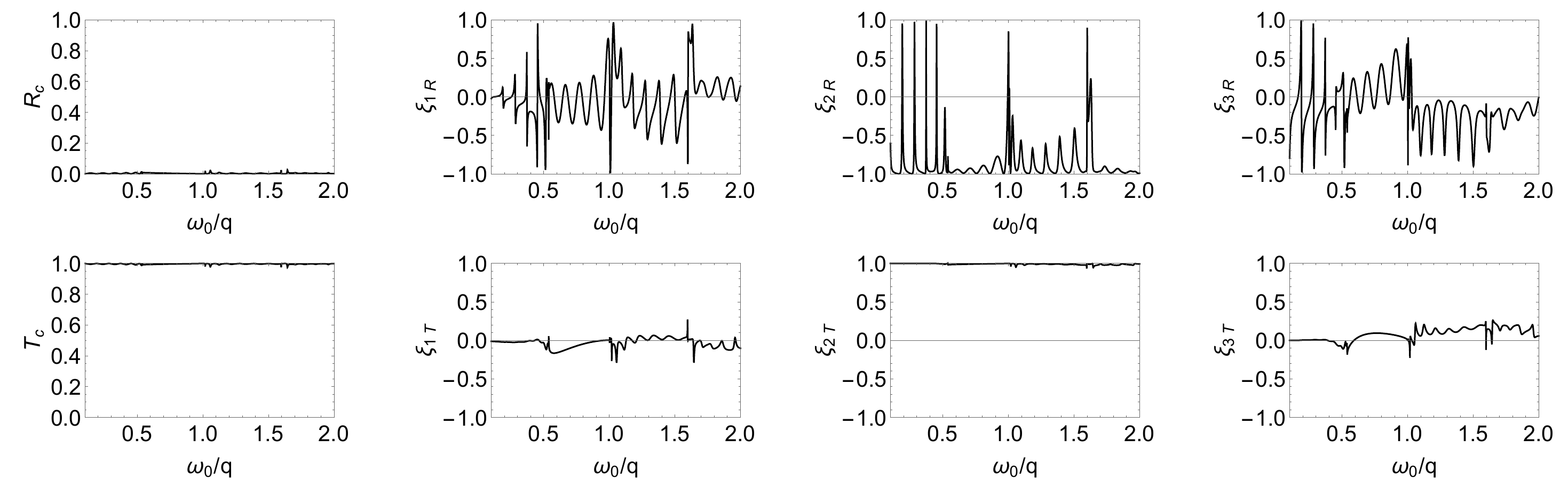}
	\includegraphics*[width=0.97\linewidth]{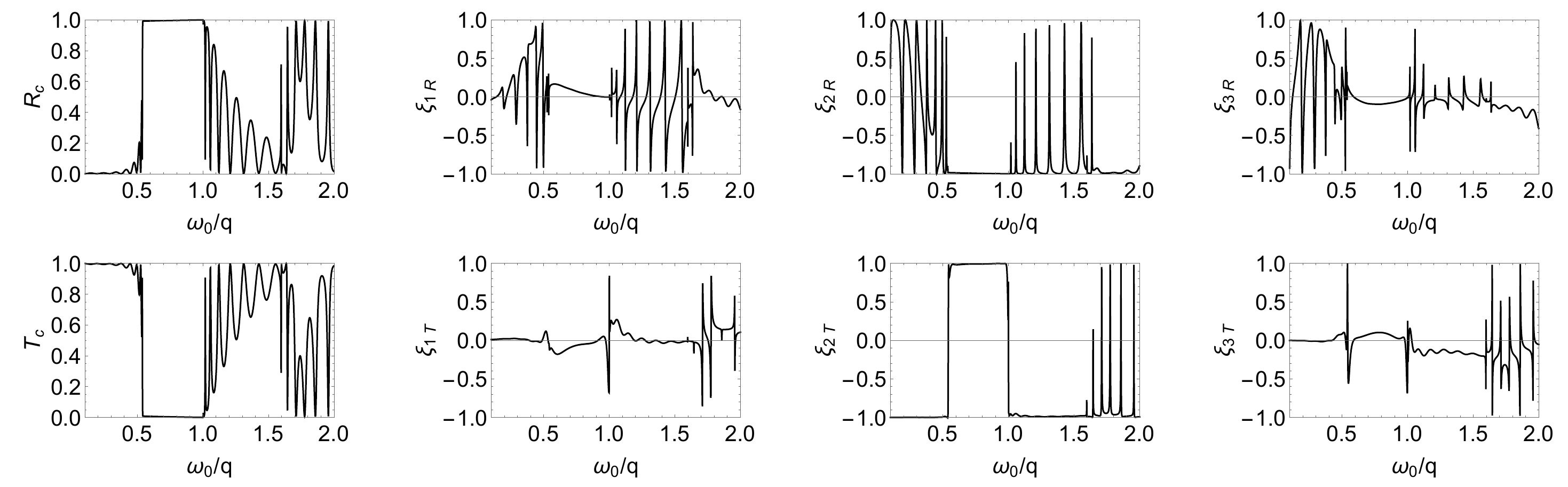}
	\caption{{\footnotesize The same as in Fig. \ref{Scattering1} but the parameters are taken as follows: $\alpha=0.28\pi$, $v=2.65$, and $\omega_p=1.34$. The dispersion law of the electromagnetic field modes is shown in Fig. \ref{DispBothFlat}. For these parameters, the first and second branches of the dispersion law are flat near the boundary of the chiral forbidden band. In the scattering data, this leads to a wide energy range where the reflection coefficient of waves with left-hand polarization is equal to unity and the boundary of this region has a sharp edge.}}
	\label{Scattering3}
\end{figure}

The first equation in system \eqref{first_two_orders} does not determine $L_0(z)$ uniquely. It is clear that $L_0(z)$ can be multiplied by some function of $z$ and $L_0(z)$ will remain a solution to the first equation in \eqref{first_two_orders}. The expression for $L_0(z)$ is fixed uniquely up to a multiplicative constant by conditions \eqref{consist_eq}. Let $\bar{L}_0(z)$ be some fixed solution to the first equation in \eqref{first_two_orders}. Then $L_0(z)=f(z)\bar{L}_0(z)$. It follows from conditions \eqref{consist_eq} that
\begin{equation}
    |f(z)|^2=\frac{c}{2 k_3\bar{L}_0^\dag M_0\bar{L}_0 +\bar{L}_0^\dag M_1\bar{L}_0},\qquad
    (\arg f(z))'=\im\frac{2 k_3\bar{L}'^\dag_0 M_0\bar{L}_0 +\bar{L}'^\dag_0 M_1\bar{L}_0}{2 k_3\bar{L}_0^\dag M_0\bar{L}_0 +\bar{L}_0^\dag M_1\bar{L}_0},
\end{equation}
where $c>0$ is some constant.

\begin{figure}[tp]
	\begin{subfigure}{0.24\linewidth}
		\includegraphics*[width=1\linewidth]{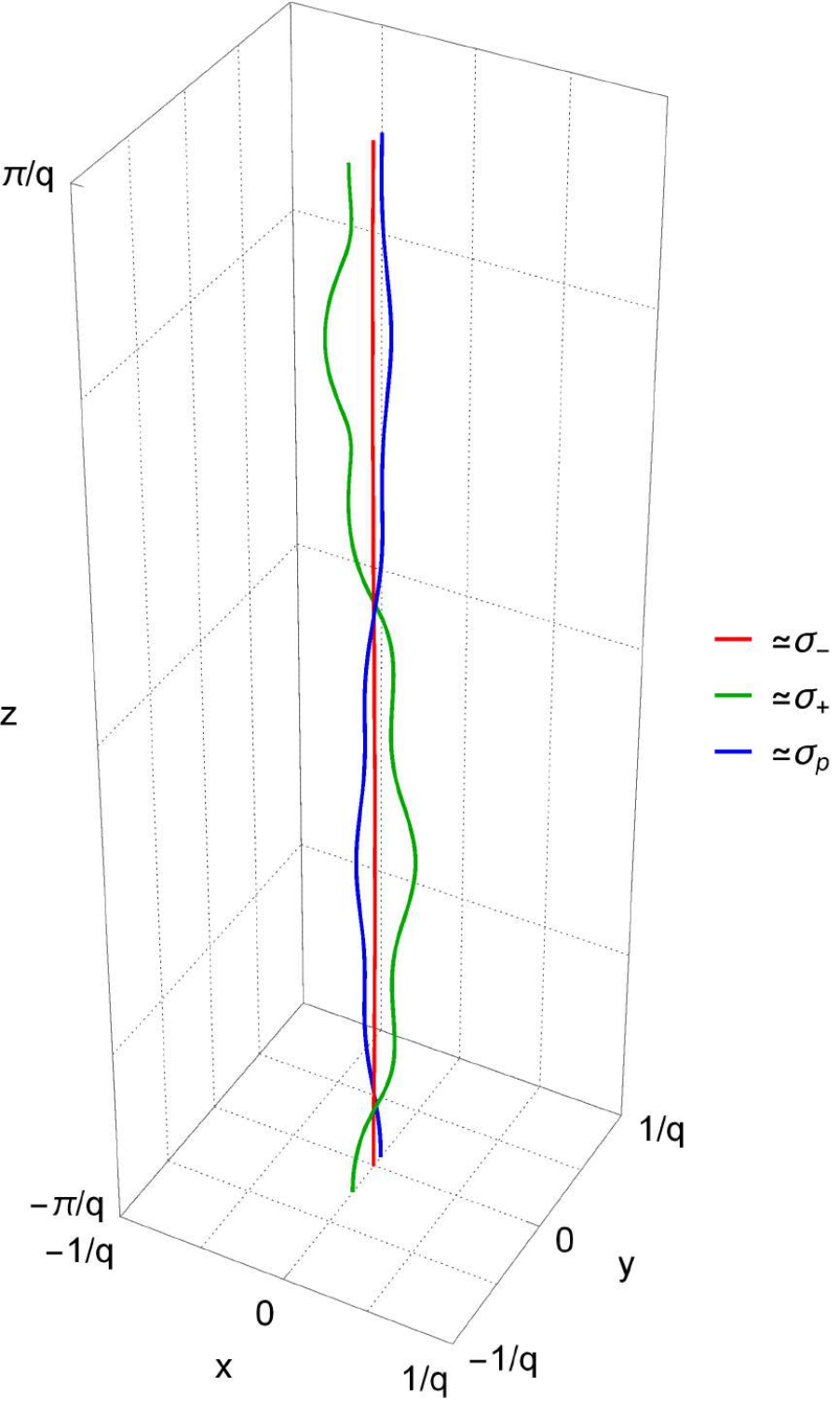}
		\caption{\label{aPoynt}}
	\end{subfigure}
	\begin{subfigure}{0.24\linewidth}
		\includegraphics*[width=1\linewidth]{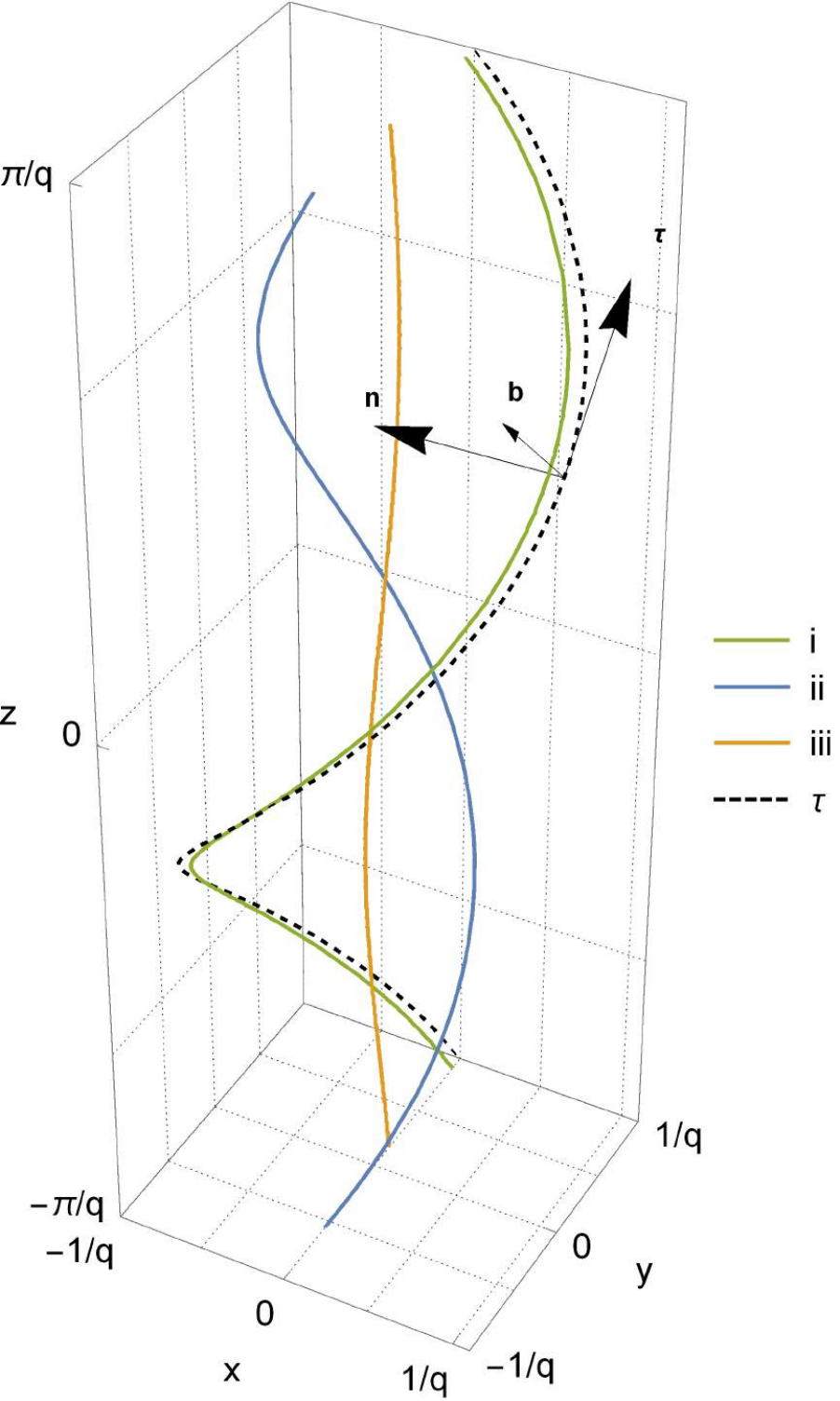}
		\caption{\label{bPoynt}}
	\end{subfigure}
	\begin{subfigure}{0.24\linewidth}
		\includegraphics*[width=1\linewidth]{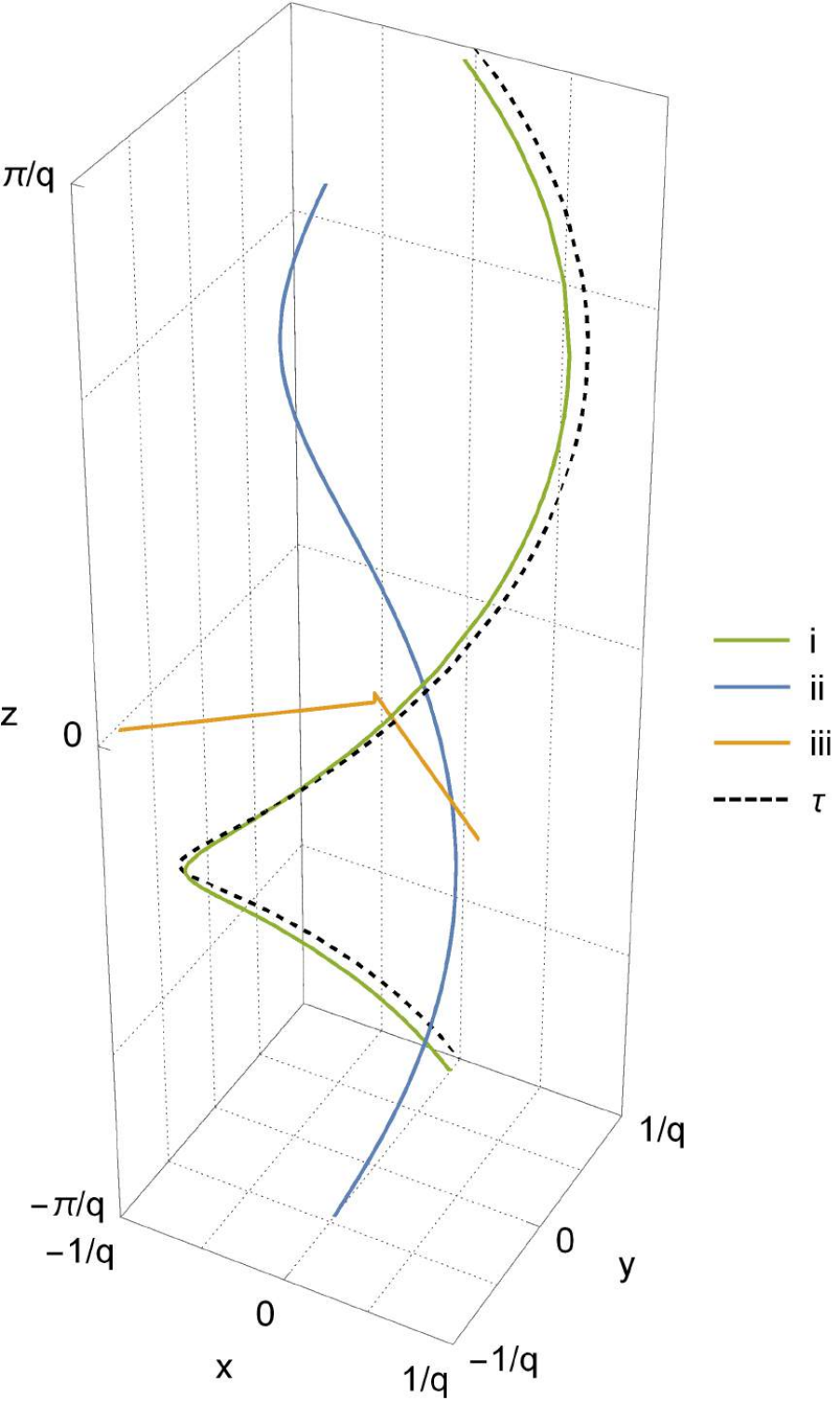}
		\caption{\label{cPoynt}}
	\end{subfigure}
	\begin{subfigure}{0.24\linewidth}
		\includegraphics*[width=1\linewidth]{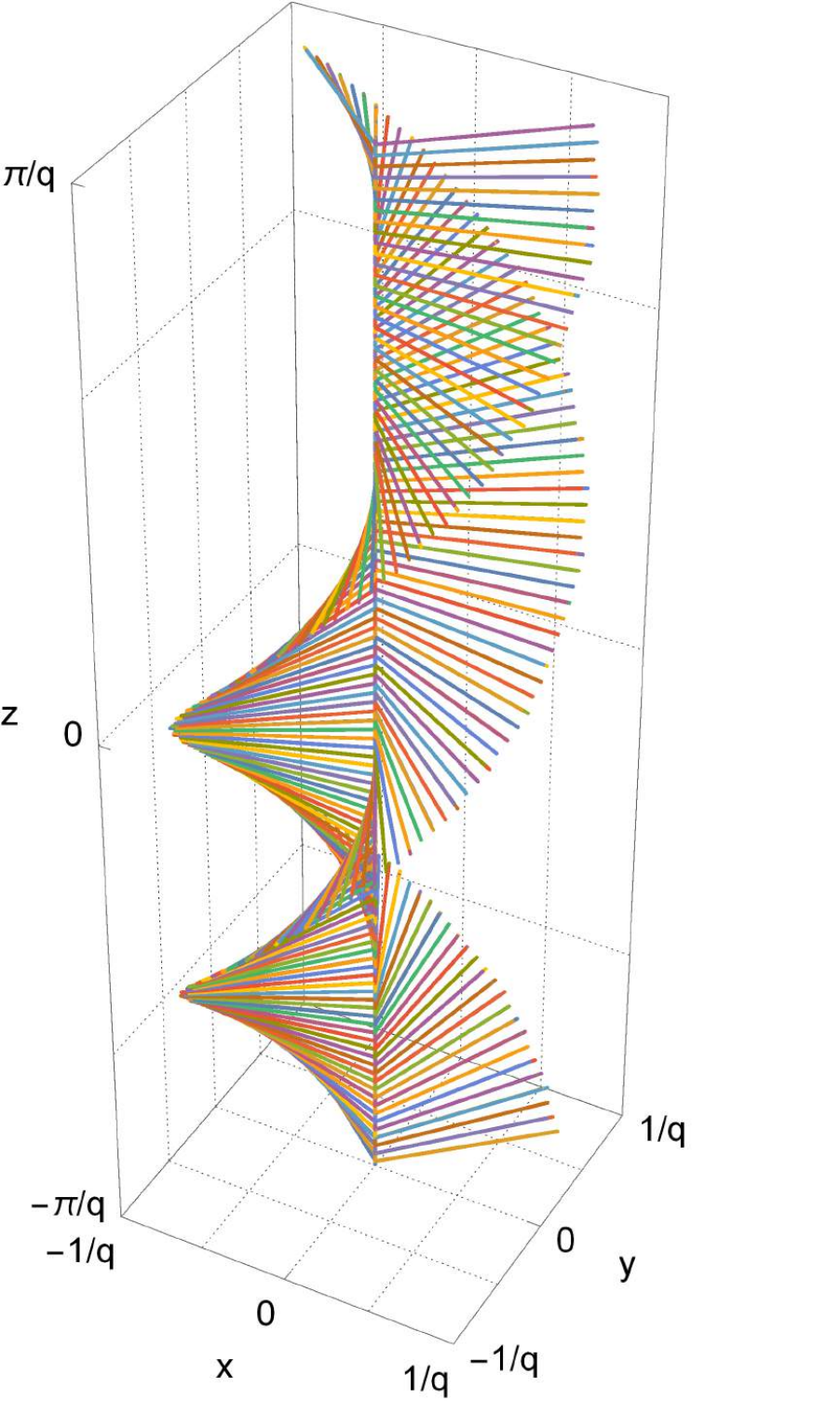}
		\caption{\label{dPoynt}}
	\end{subfigure}
	\caption{{\footnotesize The instantaneous integral curves of the Poynting vector. The plot $(a)$: The solid lines show the integral curves of the Poynting vector for photon energies close to the asymptotics $\sigma_{\pm}$, $\sigma_p$. The values of parameters are $\alpha = \pi/4$, $v=1$, $\omega_p=0.5$, $\e_h=1$, $k_3=3$, and $q=1$. The plot $(b)$: The solid lines show the integral curves of the Poynting vector for the values of parameters $\alpha = \pi/4$, $v=1$, $\omega_p=50$, $\e_h=1$, $k_3 =45$, $k_\perp=0$, and $q=1$. The dashed line represents the integral curve of the vector $\bs\tau$. The Frenet frame is shown by vectors. The numbering of the dispersion law branches is chosen in accordance with formula \eqref{WKBBranches}. The plot $(c)$: The same as in plot $(b)$. The values of parameters are $\alpha = \pi/4$, $v=1$, $\omega_p=50$, $\e_h=1$, $k_3=55$, $k_\perp=0$, and $q= 1$. The integral curve of one of the dispersion law branches is degenerate. The plot $(d)$: The degenerate integral curves of the Poynting vector of one of the dispersion law branches for different $z$. The values of parameters are the same as in plot $(c)$. The congruence of integral curves constitutes a helicoid.}}\label{PoyntingCurves}
\end{figure}

For the system of equations \eqref{MaxwellEq1}, the generalized eigenvalue problem \eqref{first_two_orders} possesses six different solutions. In the case $v=1$, the expressions for the eigenvalues are significantly simplified and further we will consider only this case. Then
\begin{equation}\label{WKBBranches}
    i)\;k_3=\pm\sqrt{\bar{k}_3^2-\omega_p^2},\qquad ii)\;k_3=\pm \bar{k}_3,\qquad iii)\; k_3=\frac{\pm \omega_0 -k_\perp\sin\al\cos\theta}{\cos\al}.
\end{equation}
Notice that the last expression can be written as $\omega_0=\pm(\bs\tau\spk)$, i.e., these modes describe the propagation of electromagnetic waves along the conducting wires. Below we will verify this fact by finding the explicit expression for the Poynting vector. Applying the procedure outlined above, we obtain the eigenvectors for the branches of the dispersion law $(i)_\pm$:
\begin{equation}\label{L1pm}
    L_0^{1\pm}=\frac{ \big\{k_\perp k_3\cos\al -(e^{i\theta}\omega_0^2-\cos\theta k_\perp^2)\sin\al,k_\perp k_3\cos\al -(e^{-i\theta}\omega_0^2-\cos\theta k_\perp^2)\sin\al,\sqrt{2}\omega_0\omega_p\big\}}{2\omega_0|k_3|^{1/2} [(k_3\sin\al-k_\perp\cos\al\cos\theta)^2+\omega_p^2+k_\perp^2\sin^2\theta]^{1/2}} e^{ik_3z}\Big|_{k_3=\pm\sqrt{\bar{k}_3^2-\omega_p^2}};
\end{equation}
for the branches $(ii)_\pm$:
\begin{equation}\label{L2pm}
    L_0^{2\pm}=\frac{ \big\{ -k_\perp\cos\al +k_3e^{i\theta}\sin\al,k_\perp\cos\al -k_3e^{-i\theta}\sin\al,0\big\}}{|k_3|^{1/2} [(k_3\sin\al-k_\perp\cos\al\cos\theta)^2+k_\perp^2\sin^2\theta]^{1/2}} e^{ik_3z}\Big|_{k_3=\pm |\bar{k}_3|};
\end{equation}
for the branch $(iii)_+$:
\begin{equation}\label{L3p}
    L_0^{3+}=\frac{ \big\{ \omega_p\cos^2\al (\omega_0\sin\al e^{i\theta} -k_\perp) ,\omega_p\cos^2\al (\omega_0\sin\al e^{-i\theta} -k_\perp),\sqrt{2} \big((k_\perp\cos\theta-\omega_0\sin\al)^2 +k_\perp^2\cos^2\al\sin^2\theta\big)\big\} \exp[i S_+(z)]}{ 2\big[\omega_0\cos\al\big((k_\perp\cos\theta-\omega_0\sin\al)^2 +k_\perp^2\cos^2\al\sin^2\theta\big) \big((k_\perp\cos\theta-\omega_0\sin\al)^2 +k_\perp^2\cos^2\al\sin^2\theta +\omega_p^2\cos^2\al\big)\big]^{1/2}} ;
\end{equation}
and for the branch $(iii)_-$:
\begin{equation}\label{L3m}
    L_0^{3-}=\frac{ \big\{ \omega_p\cos^2\al (\omega_0\sin\al e^{i\theta} +k_\perp) ,\omega_p\cos^2\al (\omega_0\sin\al e^{-i\theta} +k_\perp),\sqrt{2} \big((k_\perp\cos\theta+\omega_0\sin\al)^2 +k_\perp^2\cos^2\al\sin^2\theta\big)\big\} \exp[i S_-(z)]}{ 2\big[\omega_0\cos\al\big((k_\perp\cos\theta+\omega_0\sin\al)^2 +k_\perp^2\cos^2\al\sin^2\theta\big) \big((k_\perp\cos\theta+\omega_0\sin\al)^2 +k_\perp^2\cos^2\al\sin^2\theta +\omega_p^2\cos^2\al\big)\big]^{1/2}} ;
\end{equation}
where
\begin{equation}
    S_\pm(z)=\frac{\pm \omega_0z-\sin\al\sin\theta k_\perp/q}{\cos\al}.
\end{equation}
One can see that there are no nontrivial turning points in this case.

The Stokes parameters characterizing the polarization of an electromagnetic wave are obtained by using formulas \eqref{Stokes_param}. It is not difficult to find these parameters for the above solutions in the paraxial limit $k_\perp\rightarrow0$. In this case,
\begin{equation}\label{Stokes_param_WKB}
    i)\;\xi_2=0,\quad\xi_3+i\xi_1=e^{2iqz};\qquad ii)\;\xi_2=0,\quad\xi_3+i\xi_1=-e^{2iqz};\qquad iii)\;\xi_2=0,\quad \xi_3+i\xi_1=e^{2iqz}.
\end{equation}
The Stokes parameters do not depend on the choice of sign (``$\pm$'') of the branch. We see that the polarization of all the modes is linear in the approximation considered. The polarization vector of the modes $(i)$ and $(iii)$ is parallel to the vector $\mathbf{d}$, and the polarization vector of the modes $(ii)$ is orthogonal to $\mathbf{d}$. Such a helical medium can be used to rotate the plane of polarization of electromagnetic waves passing through it in the same way as, for example, in cholesterics \cite{deGennes1993,Belyakov2019,Belyakov1982}.

In Fig. \ref{Scattering2}, we present the results of numerical simulations of scattering data in the shortwave paraxial regime and, in Fig. \ref{DispWKB}, the plots of the dispersion law and of the degree of circular polarization of the corresponding modes are given. For $\omega_0<\omega_p$, only the electromagnetic waves corresponding to the branches $(ii)$ and $(iii)$ of the dispersion law propagate in the helical wired medium. If, in addition, the incident wave has the linear polarization with Stokes parameter $\xi_3=1$, then, as seen from \eqref{Stokes_param_WKB}, such a wave excites mainly the mode $(iii)$. In Figs. \ref{DispWKB}, \ref{Scattering2}, the case $\al\approx\pi/2$ is presented. In this case, the contribution of the mode $(iii)$ is suppressed since this mode has a very large momentum, $k_3^2\approx\omega_0^2/(v^2(\al-\ pi/2)^2)$, for $\omega_0\gtrsim|q|$. As a result, as seen from the first two lines in Fig. \ref{Scattering2}, in the energy range $\omega_0\in(|q|,\omega_p)$, there is a total reflection of the electromagnetic waves with $\xi_3=1$ from the helical wired medium with the exception of a series of resonances. These resonances stem from the interference of the mode $(iii)$ with the small admixture of the mode $(ii)$, which only approximately has the linear polarization with $\xi_3=-1$ at $z=0$ in this energy domain (see Fig. \ref{DispWKB}). If the incident electromagnetic wave possesses the linear polarization with $\xi_3=-1$, then the mode $(iii)$ is only excited. This mode propagates freely through the medium and, at the exit from the wired medium, it has the linear polarization with the Stokes parameters given in the second formula in \eqref{Stokes_param_WKB} with $z=L$. Therefore, in the regime we consider, the helical wired medium behaves as a linear polarization filter rotating the plane of linear polarization of the transmitted wave by an angle of $qz$.

Notice that if we consider the shortwave approximation immediately in the system of equations \eqref{MaxwellEq0}, i.e., neglect noncommutativity and replace $\hat{\spk}\rightarrow\spk$, then by combining the equations and taking into account that $v=1$, it is no difficult to deduce
\begin{equation}
    (\omega_0^2-\spk^2-\omega_p^2)(\bs\tau\mathbf{A})=0.
\end{equation}
This means that, for the modes $(ii)$ and $(iii)$, the vector $\mathbf{A}=\mathbf{E}/ik_0$ is orthogonal to the vector $\bs\tau$. The fulfillment of this property for the modes $(ii)$ and $(iii)$ can be verified directly by using explicit expressions \eqref{L2pm}, \eqref{L3p}, \eqref{L3m} and finding the component $A_3$ from relation \eqref{A_3}.

Let us obtain the direction of the Poynting vector, $\bs\Pi$, for the modes given above. Using formulas \eqref{Point_defn}, we arrive at
\begin{equation}
    ii)\; \bs\Pi\sim \spk,\qquad iii)\;\bs\Pi\sim \bs\tau.
\end{equation}
As for the modes $(i)$, in the general case the expression for the direction of the Poynting vector looks rather huge
\begin{equation}\label{Point_wkb_1}
\begin{split}
    \bs\Pi\sim\,& \big[k_\perp\sin\al\cos\theta\big(k_\perp^2(1-\sin^2\al\cos^2\theta)-k_3^2(3\cos^2\al-1)\big) +k_3\cos\al\big(k_3^2\sin^2\al+\\
    &+k_\perp^2(1-3\sin^2\al\cos^2\theta)\big) \big]\bs\tau -k_\perp\sin\theta\big[\omega_0^2-(k_3\cos\al+k_\perp\sin\al\cos\theta)^2\big]\mathbf{n}+\\
    &+\big[k_3\sin\al\big(\omega_0^2-k_3^2\cos^2\al
    +k_\perp^2\cos^2\theta(3\cos^2\al-1)\big) -k_\perp\cos\al\cos\theta\times\\ &\times(\omega_0^2+k_3^2(3\sin^2\al-1)-k_\perp^2\sin^2\al\cos^2\theta) \big]\mathbf{b}.
\end{split}
\end{equation}
However, in the paraxial limit, $k_\perp\rightarrow0$, this expression is simplified considerably and reduced to
\begin{equation}
    i)\;\bs\Pi\sim \omega_0^2\sin\al\spe_3 -\omega_p^2\cos\al\mathbf{d}.
\end{equation}
Thus, under the assumptions above, the integral curves of the Poynting vectors for the modes $(i)$ and $(iii)$ form helices with pitch $2\pi/|q|$. The integral curves of the Poynting vector for the modes $(ii)$ are the same as for a plane wave in vacuum (see Fig. \ref{PoyntingCurves}). In the paraxial regime, the expressions for the direction of the Poynting vector coincide with the expressions \eqref{Point_wkbpar1}, \eqref{Point_wkbpar23} obtained in the previous section using the exact solution of the Maxwell equations in the paraxial limit in the shortwave approximation.

\section{Conclusion}

Let us sum up the results. We have investigated the propagation of electromagnetic waves in helical media with strong spatial dispersion. We have found the general form \eqref{eps_hel_gen} of the permittivity tensor possessing a helical symmetry. Then we have particularized it to the case of the helical medium consisting of conducting spiral wires. Based on the symmetry reasonings, we have found the general form \eqref{diel_permit} for the permittivity tensor with spatial dispersion for such a medium in the leading order in derivatives. The presence of a strong spatial dispersion revealing as the pole in the permittivity tensor has given rise to the appearance of the additional degree of freedom -- the field of plasmons propagating along the conducting wires in the wired medium. Introducing this field, we have rewritten the Maxwell equations in the form free from singularities and spatial nonlocalities. The resulting system of equations \eqref{MaxwellEq2} appears to be rather complicated, and we have not found its general solution in a closed form. However, it admits an analytical treatment in the case of paraxial propagation of the electromagnetic waves and in the case of short wavelengths. Then we have proceeded with the analytical study of these two regimes and have verified the analytical results by numerical simulations of scattering of the electromagnetic waves by the plate made of this wired medium.

In the paraxial approximation, we have obtained the exact expression \eqref{ParaxSol} for the modes of electromagnetic field and the dispersion law \eqref{ParaxDisp} for them in the helical wired medium. It turns out that the dispersion law possesses the six branches (the three positive energy branches) and the polarization dependent forbidden bands. We have described the properties of these forbidden bands and found, in particular, the restrictions on the parameters of the helical wired medium when the band edges are flat. The photon density of states rapidly increases near these edges and the group velocity of the modes of electromagnetic field is close to zero.

The widths of the forbidden gaps are tunable. When the helix angle, $\al$, is not close to $\pi/2$ and the plasma frequency $\omega_p\gtrsim|q|$, there is only one forbidden band and it is chiral. Its width is of order $|q|$, where $2\pi/|q|$ is the pitch of the helical structure. The energy, where this forbidden band is realized, is also of order $|q|$ (see Figs. \ref{DispColored},\ref{Scattering2-5} and \ref{DispBothFlat}, \ref{Scattering3}). We have described the polarization properties of the modes of electromagnetic field and have found the corresponding Stokes parameters (see Figs. \ref{DispAsymp}-\ref{DispBothFlat}). For the energies belonging to the chiral forbidden band, the allowed modes possess a high degree of circular polarization: the mode of electromagnetic field propagating in the positive direction along the $z$ axis has the chirality $\sgn(q)$, whereas the mode of electromagnetic field propagating in the negative direction along the $z$ axis has the chirality $-\sgn(q)$. If the plasma frequency $\omega_p\ll|q|$, then the two chiral forbidden bands are realized at the energies $\omega_0\approx\omega_p$ and $\omega_0\in(|q|,\sqrt{q^2+\omega_p^2\sin^2\al})$ (see Fig. \ref{DispAsymp}, \ref{Scattering1}).

In the case when $\al\approx\pi/2$, there are also two forbidden bands, in general. For low photon energies, $\omega_0\in(0,\omega_{0m})$, where $\omega_{0m}$ is given in \eqref{omega_0m}, the total forbidden band arises (see Figs. \ref{Disp1Flat}, \ref{Scattering2-6}). If, additionally, the plasma frequency $\omega_p\gg|q|$, then the forbidden band for a linearly polarized electromagnetic wave is present at $\omega_0\in(|q|,\omega_p)$ (see Figs. \ref{DispWKB}, \ref{Scattering2}). In this interval of energies, the electromagnetic waves with only one linear polarization are transmitted through the medium. The polarization vector of these waves is parallel to the vector $\mathbf{d}\approx\bs\tau$, where $\bs\tau$ is tangent to the conducting wires \eqref{xi_tau}. The polarization plane of these waves is rotated by an angle of $qz$ in the helical wired medium, as the vector $\bs\tau$ does. Therefore, on escaping from the medium, these waves are linearly polarized with the polarization plane rotated by an angle of $qL$ and the polarization vector parallel to the vector $\mathbf{d}$.

In order to elucidate the structure of energy fluxes in such a medium, we have found the Poynting vectors for the different modes in the paraxial regime. It turns out that, in this regime, the time averaged Poynting vector for all the modes has a vanishing component along the normal to the helix describing the conducting wires. The integral curves of the time averaged Poynting vector represent helices revolving around the $z$ axis with the pitch $2\pi/|q|$ and the chirality $\sgn(q)$. As for the integral curves of the instantaneous Poynting vector, there are two types of them (see Fig. \ref{PoyntingCurves}). The integral curve of the first type has the form of a helix winding around another helix with ``frequency'' $2k_3$ in changing the variable $z$. The reference helix around which the other helix is revolving is the helix that winds around the $z$ axis with the pitch $2\pi/|q|$ and the chirality $\sgn(q)$. In the shortwave regime, the beats with the frequency $2k_3$ in the Poynting vector becomes negligible and so the integral curve of the Poynting vector tends to the corresponding reference helix. The integral curve of the second type is realized for the photon energies \eqref{energy_range} and consists of two straight lines smoothly connected by an arc. The straight lines lie in the planes $z=z_1$ and $z=z_2$, where $z_{1,2}$ are certain constants, and the length of the arc is of order $1/k_3$ (see Fig. \ref{PoyntingCurves}).

In the shortwave approximation, we have derived the explicit expressions \eqref{L1pm}-\eqref{L3m} for the modes of electromagnetic field in the helical wired medium. The paraxial approximation has not been employed. The polarization properties of these modes are rather complicated but in the paraxial regime all the six modes are linearly polarized with the Stokes parameters \eqref{Stokes_param_WKB}. Four of these modes possess the polarization vector parallel to the vector $\mathbf{d}$ and the rest two ones have the polarization vector perpendicular to $\mathbf{d}$. We have also analyzed the integral curves of the Poynting vectors for the modes \eqref{L1pm}-\eqref{L3m}. It turns out that, in the shortwave approximation, the two modes transfer the energy along the wavevector $\spk$ as in vacuum and the other two transfer it along the vector $\bs\tau$. As for the rest two modes, the explicit expression for the direction of the corresponding Poynting vector is rather huge and is given in \eqref{Point_wkb_1}. In the paraxial regime, the integral curve of the Poynting vector for these two modes takes the form of a helix revolving around the $z$ axis with the pitch $2\pi/|q|$ and the chirality $\sgn(q)$. Of course, the paraxial limit of the Poynting vectors for these shortwave modes coincides with the shortwave approximation for the Poynting vectors evaluated in the paraxial limit \eqref{Point_wkbpar1}, \eqref{Point_wkbpar23} described above.

The helical metamaterial we study in the present paper can be constructed by various means \cite{Kashke2016,Yang2012,Gansel2012,Liao2014,Li2015,Kashke2015,Jen2015,Venkataramanababu2018,Wu2011,Wu2010,LiWongChan2015}. Apart from the peculiar polarization properties of the reflected and transmitted electromagnetic waves, the helical metamaterials can be employed for generation of twisted photons \cite{Marucci2006,Karimi2009,Fang2018,Gong2018,KK2022,Barboza2015,Li2021}. The angular momentum ``freezed'' in the medium with helical symmetry is transferred to the electromagnetic wave scattered by this medium. Moreover, the photon acquiring the additional projection of the total angular momentum can be virtual and be generated by a charged particle traversing the helical metamaterial. We leave the detailed investigation of these processes for a future research. Notice that the Vavilov-Cherenkov radiation of planewave photons in a medium composed of parallel straight conducting wires was considered
in \cite{Vorobyev2012}.

\appendix
\section{Additional boundary conditions}\label{Bound_Cond_Ap}

Let us consider the smooth interface $\Si$ between the helical wired medium with permittivity \eqref{diel_permit} and the homogeneous isotropic dielectric with permittivity $\e_0$. Denote as $\mathbf{n}$ the unit vector of the normal to this surface. Then the standard boundary conditions \cite{LandLifshECM,RyazanovB} are satisfied
\begin{equation}\label{A_st_BoundCond}
	[\mathbf{A}_\perp]_{\Si}=0, \qquad [(\rot\mathbf{A})_\perp]_{\Si}=0
\end{equation}
for the components of the vector potential perpendicular to $\mathbf{n}$ and its curl.

Let $\mathbf{A}$ and $\Psi$ be the electromagnetic and plasmonic fields in the helical wired medium obeying equations \eqref{MaxwellEq0} and $\mathbf{A}^{\e}$ be the electromagnetic field in a homogeneous isotropic dielectric with permittivity $\e_0$ described by the Maxwell equations
\begin{equation}\label{MaxEqDiel}
    (\rot^2_{ij} -k_0^2\e_0\de_{ij})A_j^{\e}=0.
\end{equation}
Convolving this equation and the second equation in \eqref{MaxwellEq0} with $\mathbf{n}$, we obtain
\begin{equation}\label{n_contr_eqs}
    \e_0 k_0^2A^\e_n-(\rot\mathbf{H}^\e)_n=0,\qquad \e^{1/2}_h k_0 \tau_n\Psi+\e_h k_0^2A_n-(\rot\mathbf{H})_n=0,
\end{equation}
where $\mathbf{H}:=\rot\mathbf{A}$. Employing the decomposition
\begin{equation}
    H_i=n_i H_n+H^\perp_i,\qquad\partial_i=n_i\partial_n+\partial^\perp_i,
\end{equation}
we deduce
\begin{equation}\label{n_rotn}
    (\rot\mathbf{H})_n=(\rot\mathbf{n})_n H_n +n_i\e_{ijk}\partial_j^\perp H^\perp_k.
\end{equation}
Inasmuch as the vector $n_i$ is the normal to a smooth surface, it can be represented as $n_i=f\partial_ih$, where $f(x)$ and $h(x)$ are smooth functions. Hence, the first term on the right-hand side of \eqref{n_rotn} is zero. In virtue of the second boundary condition in \eqref{A_st_BoundCond}, we have
\begin{equation}
    [(\rot\mathbf{H})_n]_{\Si}=[n_i\e_{ijk}\partial_j^\perp H^\perp_k]_{\Si}=0.
\end{equation}
Then taking \eqref{n_contr_eqs} on the interface $\Si$, we come to
\begin{equation}
    \e^{1/2}_h k_0 \tau_n\Psi= k_0^2(\e_0 A^\e_n -\e_h A_n).
\end{equation}
Hence we see that, for $\tau_n\neq0$, the standard boundary condition by Pekar \cite{Pekar1958,Pekar21958},
\begin{equation}
    \Psi|_\Si=0,
\end{equation}
is equivalent to the condition
\begin{equation}
    [\e A_n]_\Si=0.
\end{equation}
The last additional boundary condition was used in the paper \cite{Silverinha2006}.

\end{document}